\documentclass[1p, final]{elsarticle}

\usepackage{subfig}
\usepackage{amsmath,bm}
\usepackage{color}
\usepackage{enumerate}
\usepackage{enumitem}
\usepackage{multirow}
\usepackage{amssymb}
\usepackage{epstopdf}
\usepackage{epsfig}
\usepackage[table,xcdraw]{xcolor}
\usepackage{caption}
\usepackage[linesnumbered,ruled,vlined]{algorithm2e} 

\usepackage{tikz}
\usepackage{graphicx}
\usetikzlibrary{positioning}

\usepackage{booktabs}
\usepackage{placeins}

\usepackage{bigstrut}
\usepackage{lineno,hyperref}
\modulolinenumbers[5]

\journal{ }

\begin{document}
	
\begin{frontmatter}
	
\title{A Structure-Preserving Graph Neural Solver for Parametric Hyperbolic Conservation Laws}

\author{Jiamin Jiang}
\ead{jiaminjiang@ustc.edu.cn}

\address{School of Artificial Intelligence and Data Science and Suzhou Institute for Advanced Research, University of Science and Technology of China, and Suzhou Big Data \& AI Research and Engineering Center, China}

\author{Shanglin Lv}
\ead{lvshanglin@mail.ustc.edu.cn}

\address{School of Mathematical Sciences, University of Science and Technology of China, China}

\author{Jingrun Chen}
\ead{jingrunchen@ustc.edu.cn}

\address{School of Mathematical Sciences and Suzhou Institute for Advanced Research, University of Science and Technology of China, and Suzhou Big Data \& AI Research and Engineering Center, China}

\begin{abstract}
	
Hyperbolic conservation laws govern a wide range of transport-driven dynamics featuring shocks, contact discontinuities, and complex wave interactions, posing distinct challenges for deep-learning-based surrogate modeling. While classical numerical methods provide robust and physically admissible solutions, their computational cost restricts applicability in many-query tasks such as parametric studies and design optimization. Conversely, existing neural surrogates offer rapid inference but often fail to respect intrinsic PDE structures, leading to non-physical artifacts, rollout instability, and poor generalization.

We present an interpretable, structure-preserving graph neural solver that bridges classical numerical principles with graph neural networks (GNNs). The network is designed as a learned reconstruction-and-flux operator rather than a black-box state updater, thereby inherently preserving key properties such as local conservation and upwinding. Inspired by Arbitrary high-order DERivatives schemes, we further recast message-passing GNNs as high-order space-time predictors, enabling conservative and stable neural updates with large time steps.

Evaluation is performed on challenging supersonic flow benchmarks spanning broad parametric variations in geometry, initial/boundary conditions, and flow regimes. The neural solver achieves superior long-horizon rollout stability and accuracy compared with strong surrogate baselines, outperforms low-order discretizations, and delivers orders-of-magnitude runtime speedups over high-resolution simulations.

\end{abstract}

\end{frontmatter}


\section{Introduction}

Partial differential equations (PDEs) underpin predictive modeling of complex physical phenomena across a wide range of scientific and engineering disciplines. High‑fidelity numerical solvers—built on well‑established finite‑volume (FV) and finite‑element (FE) discretizations—have reached a remarkable level of maturity and reliability. Yet their considerable computational expense often becomes prohibitive in settings that require thousands of repeated forward simulations, such as large‑scale parametric analysis, design optimization, and real‑time decision support. The computational cost of traditional solvers has spurred growing interest in data-driven and hybrid data–physics surrogates, where deep neural networks are trained to approximate evolution operators from simulation data, thereby shifting the bulk of expense to an offline training phase (e.g., \cite{li2020fourier,lu2021learning,cao2024laplace,liu2024neural,zappala2024learning,peyvan2024riemannonets}). While such models enable rapid inference and show potential for generalizing across families of PDE parameters, most demonstrated successes to date have been confined to regimes with smooth solutions or weakly nonlinear dynamics.

By contrast, PDEs admitting discontinuous solutions—most notably hyperbolic conservation laws—remain significantly underexplored, with existing studies typically limited in scope, dimensionality, or physical complexity. A central difficulty stems from the intrinsic mathematical and physical nature of hyperbolic systems. Consider, for example, the multidimensional Euler equations, whose solutions display rich nonlinear wave phenomena—shocks, contact discontinuities, expansion fans, and complex shock–shock or shock–wall interactions. In such regimes, numerical accuracy and stability depend not only on representational capacity but, more fundamentally, on adherence to underlying structural constraints: conservation, upwinding, correct wave propagation, and solution admissibility (e.g., positivity and entropy conditions). Black-box neural surrogates lack explicit mechanisms to enforce these properties and may therefore produce non-physical states, smear discontinuities, or develop spurious oscillations. More critically, when applied autoregressively, one-step errors can rapidly accumulate and amplify, leading to temporal drift and eventual rollout failure.

Over the past decades, robust high‑resolution schemes have been developed within the Godunov paradigm \cite{godunov1959finite}, where local Riemann problems at cell interfaces yield upwind numerical fluxes that enforce conservation and propagate information along characteristics. Modern Godunov-type methods pair (approximate) Riemann solvers \cite{toro2013riemann} with non–oscillatory high-order nonlinear reconstructions (e.g., MUSCL~\cite{van1997towards}, WENO~\cite{liu1994weighted,jiang1996efficient}) and strong-stability-preserving time integrators \cite{shu1988efficient,gottlieb2001strong}, enabling sharp shock capturing while controlling spurious oscillations. Their robustness stems precisely from embedding key structures directly into the algorithm. This offers a clear conceptual blueprint for reliable surrogate modeling of hyperbolic conservation laws; however, within a learning framework, how to incorporate such structure‑preserving mechanisms of classical algorithms remains open and challenging.

In this work, we introduce an interpretable and structure‑preserving graph neural solver that systematically integrates the rigor of classical Godunov-type methods with the expressive capacity of graph neural networks (GNNs). We demonstrate that the successive, geometry‑aware message‑passing and neighborhood aggregation in GNNs effectively emulate the action of wide spatial stencils used in high‑order nonlinear reconstructions, with directional edge‑based propagation naturally inducing one‑sided biasing in the latent representations. Crucially, the network is not treated as a black‑box state updater, but rather as a \textit{learned reconstruction‑and‑flux operator}: the GNN outputs interface‑oriented states via edge‑wise decoding, a differentiable Riemann solver maps them to consistent numerical fluxes, and a conservative flux‑aggregation layer advances the solution. This design enforces local conservation and upwinding directly at the architectural level.

A further limitation is the reliance of high-fidelity solvers on explicit time integration constrained by the Courant–Friedrichs–Lewy (CFL) condition, leading to prohibitively small time steps and densely sampled trajectories. To circumvent this, we propose a formulation inspired by Arbitrary high-order DERivatives (ADER) schemes \cite{titarev2002ader}, wherein the message-passing GNN serves as a \textit{high-order space-time predictor} that encodes fine-scale dynamical evolution over each coarse time interval. The resulting implicit‑like one‑step update remains fully conservative and compatible with Riemann solvers, yet operates stably with $\Delta t \gg \Delta t_{\mathrm{CFL}}$, dramatically improving both training efficiency and inference speedups.

To evaluate the proposed framework, we conduct extensive experiments on four canonical families of supersonic flow benchmarks, generated using a high‑order discontinuous Galerkin spectral element (DGSEM) solver on unstructured meshes. The datasets encompass wide parametric variations in geometry and free‑stream Mach number, covering diverse initial/boundary conditions and flow regimes. Across all benchmarks, our neural solver demonstrates significantly improved long‑horizon rollout stability, accuracy (with up to 80\% error reduction), and qualitative reliability compared to a strong GNN-based surrogate baseline. It also outperforms low-order DGSEM discretizations. These advantages are especially pronounced in capturing sharp shock fronts, contact discontinuities, and coupled shock–subsonic interactions, while suppressing non‑physical artifacts and curbing error accumulation during autoregressive inference. Furthermore, the solver achieves substantial runtime acceleration—exceeding two orders of magnitude—relative to high‑resolution reference simulations.

\section{Related Work}

\subsection{ML-Enhanced Numerical Methods}

A prominent line of research seeks to enhance classical numerical solvers through machine‑learning (ML) techniques. In this hybrid paradigm, neural networks are integrated to augment or replace specific algorithmic steps—retaining the foundational strengths of established discretizations while leveraging data‑driven modules to improve accuracy or computational efficiency. Representative efforts include learning discontinuity detectors or ``troubled‑cell” sensors that refine shock‑capturing capability \cite{sun2020convolution,feng2021characteristic}, as well as enhancing high-resolution schemes (e.g., WENO) by adjusting/optimizing smoothness indicators, nonlinear weights, or reconstruction coefficients \cite{kossaczka2021enhanced,nogueira2024machine}. Such learned components can sharpen discontinuities while preserving high‑order accuracy in smooth regions; in certain designs, they act directly as data‑driven weighting functions within compact stencils \cite{bezgin2022weno3}. Another strategy involves learning the numerical flux function itself within the FV framework. Chen et al. \cite{chen2024learning} construct data-driven interface fluxes for unknown systems and show that neural updates formulated via conservative flux differences yield correct shock-dominated dynamics. Similarly, Morand et al. \cite{morand2024deep} optimize numerical fluxes from Riemann-problem solutions using a three-point stencil, achieving improved accuracy for scalar conservation laws over classical Godunov and Engquist-Osher fluxes. Note that both approaches are restricted to one-dimensional hyperbolic PDEs and first-order FV discretizations.

While such hybrid strategies illustrate the promise of combining ML with established numerical schemes, they largely remain confined to localized improvements—often targeting individual components rather than rethinking the end‑to‑end solution process within a coherent, structure‑preserving learning framework. Furthermore, because the replaced modules typically account for only a fraction of the total computational cost, the achievable runtime acceleration is inherently capped.

\subsection{Operator Learning and Neural Surrogates}

Recent years have witnessed rapid growth in data-driven neural surrogates for computational fluid dynamics, aiming to enable fast, end-to-end predictions of flow fields for parametric studies and real-time applications \cite{lino2023current}. Early successes were largely built upon grid-based architectures such as convolutional neural networks (CNNs) and U-Net variants \cite{sekar2019fast,bhatnagar2019prediction,thuerey2020deep}, as well as Transformers and GNNs \cite{sanchez2020learning,pfaff2020learning,brandstetter2022message,geneva2022transformers,lam2023learning,li2023scalable,wu2024transolver,jiang2025local}, which better accommodate irregular geometries and unstructured discretizations. In parallel, operator learning has become a principled framework for approximating solutions of parametric PDEs, mapping problem inputs (coefficients, boundary/initial conditions) to solution functions across families of instances. Representative approaches, including DeepONet, Fourier Neural Operators, and their variants \cite{li2020fourier,lu2021learning,jin2022mionet}, have demonstrated strong performance on a range of elliptic/parabolic problems (smooth flow regimes).

Despite these advances, accurate modeling of PDEs—especially hyperbolic conservation laws—remains fundamentally challenging and is not adequately addressed by existing black-box neural surrogates. High-fidelity solutions are constrained by intrinsic PDE structures, but standard neural architectures cannot preserve them. To alleviate these issues, data–physics hybrid strategies—physics-informed neural networks (PINNs) and related approaches—incorporate PDE residuals or physical constraints into the training objective \cite{raissi2019physics,mao2020physics,wang2021learning,coutinho2023physics,de2024wpinns,liu2024discontinuity,cassia2025godunov}. While such loss regularization can improve accuracy and generalization to some extent, it imposes only soft constraints: satisfaction is encouraged during optimization but not guaranteed at inference, and violations can still arise under distribution shift and long-horizon rollouts. This gap highlights an important yet under-explored direction in the surrogate literature: embedding conservation laws and key structures directly into the model architecture as hard constraints, so that invariants and stability mechanisms are preserved by design rather than learned implicitly.

\subsection{Conservation-Enforcing Neural Surrogates}

Physical conservation laws are fundamental invariants of many dynamical systems and PDEs, and their satisfaction is essential for numerical fidelity and long-horizon stability. Previous works broadly fall into two categories. (i) \emph{Correction-based methods} perform post-processing to network outputs by solving constrained minimization problems \cite{lee2021deep}, or by projecting predictions onto invariant manifolds \cite{cardoso2025exactly}. While these techniques can guarantee conservation, they incur extra inference overhead. Moreover, most formulations are developed for low-dimensional conserved quantities in dynamical systems, and their extension to nonlinear PDEs can be nontrivial. (ii) \emph{Architectural-based methods} embed conservation directly into network design. Examples include divergence-free parameterizations that satisfy continuity equations \cite{richter2022neural}, and divergence-free output layers with numerical differentiation incorporated into neural operators \cite{liu2023harnessing}. For Lagrangian systems, Müller \cite{muller2023exact} leveraged Noether's theorem to guarantee conservation of quantities associated with symmetry invariance. Recent graph-based surrogates establish conservation-informed message passing and treat GNNs as FV-like flux exchanges on general graphs \cite{horie2024graph}.

For shock-dominated hyperbolic systems with multiple coupled conserved variables, correct wave propagation and discontinuous solutions rely on discrete-level local conservation through interface flux balances. Despite the aforementioned advances, existing methods primarily target global/linear constraints or smooth conservation forms, and therefore remain insufficient for preserving the key numerical properties required by multidimensional hyperbolic PDEs.

\section{Finite Volume Methods}

Hyperbolic conservation laws describe a broad range of transport-driven phenomena such as compressible aerodynamics, high-speed propulsion, and astrophysical flows. This work focuses on the two-dimensional unsteady Euler equations. Let $\Omega \subset \mathbb{R}^2$ denote the computational domain with spatial coordinates $\bm{x}=(x,y)\in\Omega$. The governing equations in conservative form are written as
\begin{equation}
	\label{eq:consEqns}
	\left\{
	\begin{aligned}
		& \frac{\partial \bm{u}(\bm{x},t)}{\partial t} + \nabla \cdot \mathbf{F}(\bm{u}) = 0, \\
		& \bm{u}(\bm{x},0) = \bm{u}^0(\bm{x}),
	\end{aligned}
	\right.
\end{equation}
subject to suitable boundary conditions. Here, $\bm{u}=(\rho, \rho v_1, \rho v_2, E)^{\top}$ and $\bm{h}=(\rho, v_1, v_2, p)^{\top}$ denote the vectors of conserved and primitive variables, respectively, with the density $\rho$, velocity components $(v_1,v_2)$, and the total energy $E$. The nonlinear physical fluxes $\mathbf{F}(\bm{u})=(\mathbf{F}_1,\mathbf{F}_2)$ in the $x$- and $y$-directions are given by
\begin{equation}
	\mathbf{F}_1 =
	\begin{pmatrix}
		\rho v_1\\
		\rho v_1^2+p\\
		\rho v_1 v_2\\
		v_1 (E+p)
	\end{pmatrix},\quad
	\mathbf{F}_2 =
	\begin{pmatrix}
		\rho v_2\\
		\rho v_1 v_2\\
		\rho v_2^2+p\\
		v_2 (E+p)
	\end{pmatrix},
\end{equation}
and the system is closed by the ideal-gas equation of state
\begin{equation}
	p = (\gamma-1)\left(E-\tfrac{1}{2}\rho \, (v_1^2+v_2^2)\right)
\end{equation}
with ratio of specific heats $\gamma=1.4$.

The design of our structure-preserving graph neural solver is underpinned and guided by the core algorithmic components of finite volume (FV) methods for hyperbolic conservation laws, which we briefly outline here.

\subsection{Godunov Framework} \label{GodunovFramework}

\begin{figure}[!htb]
	\centering
	\subfloat[Region around target cell $i$]{
		\includegraphics[scale=0.36]{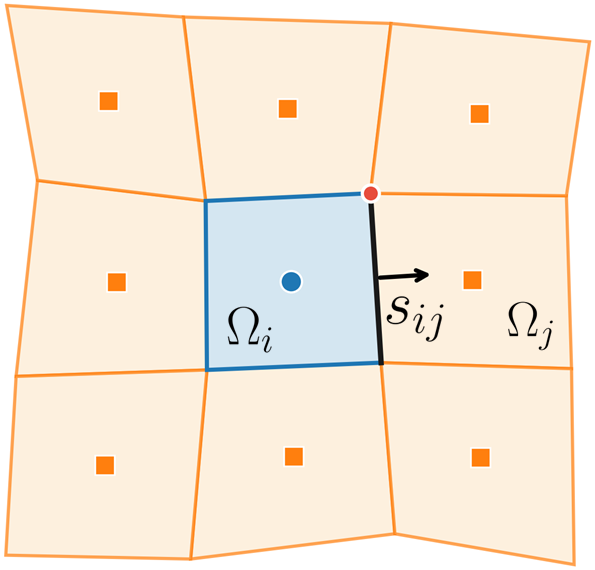}}
	\qquad
	\subfloat[Numerical flux]{
		\includegraphics[scale=0.33]{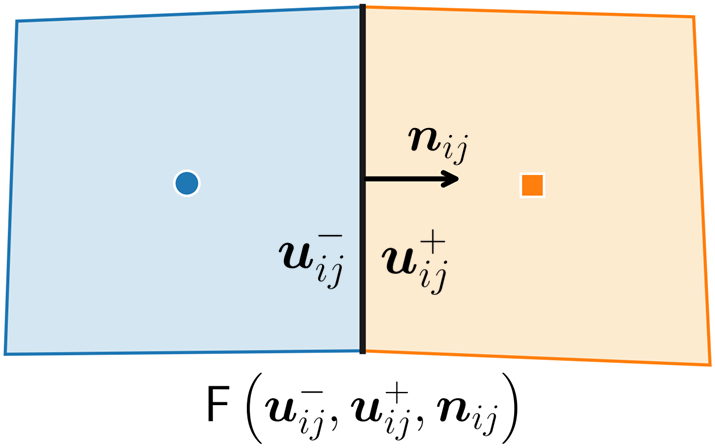}}
	\caption{Schematic illustration of the mesh ingredients and numerical flux in the Godunov framework.}
	\label{fig:FVmesh}
\end{figure}

In an FV discretization, $\Omega$ is tessellated into a set of non-overlapping control volumes (triangular or quadrilateral cells). For each cell $\Omega_i$, let $s_{ij}$ denote the shared edge (interface) between $\Omega_i$ and $\Omega_j$, where $j\in\mathcal{N}(i)$ with $\mathcal{N}(i)$ being the index set of the adjacent cells $\Omega_j$ (see \textbf{Figure~\ref{fig:FVmesh}}). The quantities $|\Omega_i|$ and $|s_{ij}|$ measure the cell area and the edge length, respectively.

The unknowns are the cell-averaged states
\begin{equation}
	\overline{\bm{u}}_i(t) = \frac{1}{|\Omega_i|} \int_{\Omega_i} \bm{u}(\bm{x},t) \, \mathrm{d}x \mathrm{d}y.
\end{equation}
Integrating Eq.~(\ref{eq:consEqns}) over $\Omega_i$ and applying the divergence theorem yields the semi-discrete form
\begin{equation}
	\label{eq:semidiscrete}
	\frac{\mathrm{d}\overline{\bm{u}}_i(t)}{\mathrm{d}t} = - \frac{1}{|\Omega_i|} \oint_{\partial \Omega_i} \mathbf{F}(\bm{u})\cdot \bm{n}\, \mathrm{d}s
\end{equation}
with $\bm{n}$ the outward unit normal along the cell boundary $\partial \Omega_i$. The boundary integral is approximated by summing interface fluxes
\begin{equation}
	\label{eq:lineIntNFlux}
	\oint_{\partial \Omega_i} \mathbf{F}(\bm{u})\cdot \bm{n}\, \mathrm{d}s \approx \sum_{j\in\mathcal{N}(i)} |s_{ij}| \; \mathsf{F} \left ( \bm{u}_{ij}^{-}, \bm{u}_{ij}^{+}, \bm{n}_{ij} \right ),
\end{equation}
where $\mathsf{F} \left ( \bm{u}^{-}, \bm{u}^{+}, \bm{n} \right )$ is a numerical flux function depending on the left and right states $\bm{u}_{ij}^{-}$ (inside the target cell) and $\bm{u}_{ij}^{+}$ (resp. outside) of the interface (see \textbf{Figure~\ref{fig:FVmesh}}).

Within the Godunov framework \cite{godunov1959finite}, the numerical flux at each cell interface is obtained by solving a local Riemann problem, in the local coordinate $\tau$ aligned with the interface normal $\bm{n}$
\begin{equation}
	\begin{aligned}
		& \frac{\partial \bm{u}(\tau, t)}{\partial t} + \frac{\partial \big ( \mathbf{F}(\bm{u})\cdot \bm{n} \big )}{\partial \tau} = 0,  \\
		& \bm{u}(\tau, 0) = 
		\begin{cases}
			\, \bm{u}^{-}, & \tau<0 \\
			\, \bm{u}^{+}, & \tau>0
		\end{cases}
	\end{aligned}
\end{equation}
with initial condition given by the jump states. The Riemann problem describes the evolution of such a discontinuity, giving rise to nonlinear waves propagating along characteristic directions. Upwind methods explicitly exploit this wave propagation information to ensure stability and physical fidelity \cite{toro2013riemann}. In practice, solving the exact Riemann problem is computationally prohibitive. Instead, approximate Riemann solvers (Rusanov, HLL-family, Roe, etc.) are widely adopted. Riemann solvers guarantee that perturbations are propagated in an upwind-consistent manner, forming the backbone of modern Godunov-type FV methods for compressible flows.

\subsection{High-order Reconstructions} \label{HighorderRecons}

High-order spatial discretizations have been developed to overcome the excessive numerical diffusion of first-order approximations. A standard approach is the piecewise linear reconstruction (second-order accuracy) as in MUSCL schemes \cite{van1979towards,van1997towards}, where the solution of a scalar variable within cell $i$ is represented as
\begin{equation}
	u(\bm{x}) = \overline{u}_i + \nabla u_i \cdot (\bm{x} - \bm{x}_i)
\end{equation}
with $\bm{x}_i$ denoting the cell centroid and $\nabla u$ the vectorial slope, often obtained by least–squares fitting of neighboring cell averages \cite{barth1989design}. This reconstruction is typically applied component–wise to either conservative or primitive variables.

Near shocks or discontinuities, however, such reconstructions generate spurious oscillations. To prevent the formation of new local extrema, nonlinear slope limiters are introduced \cite{venkatakrishnan1995convergence,park2010multi}. The limited reconstruction takes the form
\begin{equation}
	u(\bm{x}) = \overline{u}_i + \varphi_i \nabla u_i \cdot (\bm{x} - \bm{x}_i),
\end{equation}
where $\varphi_i\in[0,1]$ is a limiter coefficient. The limiting strategies ensure that any reconstructed values satisfy a discrete maximum principle for scalar problems under suitable CFL conditions, and in practice they provide non–oscillatory shock–capturing capability for the Euler system. At a given edge midpoint $\bm{x}_{ij}$, the reconstructed interface values become
\begin{equation}
	u_{ij}^{-} = \overline{u}_i + \varphi_i \nabla u_i \cdot (\bm{x}_{ij} - \bm{x}_i), 
	\qquad
	u_{ij}^{+} = \overline{u}_j + \varphi_j \nabla u_j \cdot (\bm{x}_{ij} - \bm{x}_j).
\end{equation}

To reach the third–order spatial accuracy, quadratic polynomial reconstruction can be employed, with the unlimited form as
\begin{equation}
	u(\bm{x}) = \overline{u}_i + \nabla u_i \cdot \bm{r} + \tfrac{1}{2} \, \bm{r}^{\top} \mathcal{H}_i \, \bm{r},
\end{equation}
where $\bm{r} = (\bm{x} - \bm{x}_i)$ and $\mathcal{H}_i$ is an approximation to the Hessian matrix $\nabla^2 u$. For third- and fourth-order schemes, the line integrals in Eq.~(\ref{eq:lineIntNFlux}) need to be discretized by two-point Gaussian quadrature on every edge. At each quadrature point, the reconstructed left and right states are passed to a Riemann solver for evaluating the numerical flux.

An alternative family of reconstructions is weighted essentially non-oscillatory (WENO) schemes \cite{liu1994weighted,jiang1996efficient,hu1999weighted}. The key idea of WENO is to build, for each target cell, a high-order polynomial that preserves conservation. The reconstruction is carried out on a wide spatial stencil $\left\{ \mathcal{S}_z \right\}$, which is the union of multiple small sectorial stencils. Each such stencil consists of the target cell $\Omega_i$ with several layers of neighbors. A polynomial $\mathfrak{P}_z(\bm{x})$ on $\mathcal{S}_z$ is determined so that it matches the cell averages, while ensuring that the mean over $\Omega_i$ coincides with $\overline{u}_i$. The final approximation is then formed as a convex combination of these candidate polynomials
\begin{equation}
	\mathfrak{P}(\bm{x}) = \sum_{z=1} \, \omega_z \, \mathfrak{P}_z(\bm{x}),
\end{equation}
where the nonlinear weights $\omega_z$ (satisfying $\sum_{z} \omega_z = 1$) are computed based on linear weights and smoothness indicators $\beta_z$. The smoothness of a candidate polynomial is measured by 
\begin{equation}
	\beta_z = \sum_{|\alpha|=1}^{K} |\Omega_i|^{|\alpha|-1} \int_{\Omega_i} \left( 
	\frac{\partial^{|\alpha|}}{\partial x^{\alpha_1} \partial y^{\alpha_2}} 
	\mathfrak{P}_z(\bm{x}) 
	\right)^2 \mathrm{d}x \mathrm{d}y,
\end{equation}
where $K$ is the polynomial degree and $\alpha$ is a multi-index. $\omega_z$ favors smooth stencils in regions without discontinuities, thereby recovering optimal order of accuracy, while suppressing stencils that cross discontinuities, thereby preventing oscillations. Through this mechanism, WENO schemes achieve stable and robust shock capturing without requiring explicit limiters.

On each cell interface, point-wise data are obtained from the reconstructed polynomials, which are then used for the evaluation of numerical fluxes. To advance the solution in time, the semi-discrete form (\ref{eq:semidiscrete}) can be cast as a method-of-lines ordinary differential equation (ODE) system. Time integration is typically performed with explicit schemes, among which high-order total variation diminishing (TVD) Runge–Kutta methods are a popular choice \cite{shu1988efficient,gottlieb2001strong}.

\section{Neural PDE Solvers}

Consider the problem of solving parametric Euler equations. In a FV discretization, $\Omega$ is tessellated into an unstructured mesh $\left\{ \Omega_i \right\}$ comprising $n$ cells. The discrete state vector at time level $t^m$ is denoted by $\mathbf{U}^{m} = \left\{ \bm{u}_i^m \right\}_{i=1}^{n}$, where $\bm{u}_i^m$ represents the solution value at each cell centroid. A high-fidelity numerical solver $\mathbb{H}$, based on high-order spatial and temporal schemes, maps the current mesh state $\mathbf{U}^{m}$ to the next timestep state $\mathbf{U}^{m+1}$. A trajectory (sequence) of states $\left ( \mathbf{U}^{0}, \mathbf{U}^{1}, ..., \mathbf{U}^{n_t} \right )$ can be computed by repeated application of $\mathbf{U}^{m+1} = \mathbb{H} \left ( \mathbf{U}^{m}, \bm{\mu} \right )$ across $n_t$ timesteps, where $\bm{\mu}$ denote PDE parameters (e.g., coefficients, boundary conditions, and geometries).

The deep learning task aims to replace the costly high-fidelity solver $\mathbb{H}$ with a neural solver $\mathbb{N}$ that predicts the right-hand side (state increment) of the semi-discrete system (\ref{eq:semidiscrete}). A standard choice is a first-order explicit scheme. Given the current state and a selected timestep size $\Delta t$, the discrete update formula is given as
\begin{equation} 
	\label{eq:disupdfor}
	\mathbf{U}^{m+1} \approx \widetilde{\mathbf{U}}^{m+1} = \mathbf{U}^{m} + \Delta t \; \mathbb{N} \left ( \mathbf{U}^{m}, \bm{\mu}; \Theta \right ),
\end{equation}
where $\widetilde{\mathbf{U}}^{m+1}$ denotes the predicted state from the neural solver, and $\Theta$ collects the network parameters learned through end-to-end, data-driven optimization. It is important to note that the discrete update should operate on the conservative variables $\bm{u}$, ensuring consistency with the underlying conservation laws, rather than the primitive variables $\bm{h}$.

\begin{figure}[!htb]
	\centering
	\subfloat[Training and inference workflows.]{
		\includegraphics[scale=0.32]{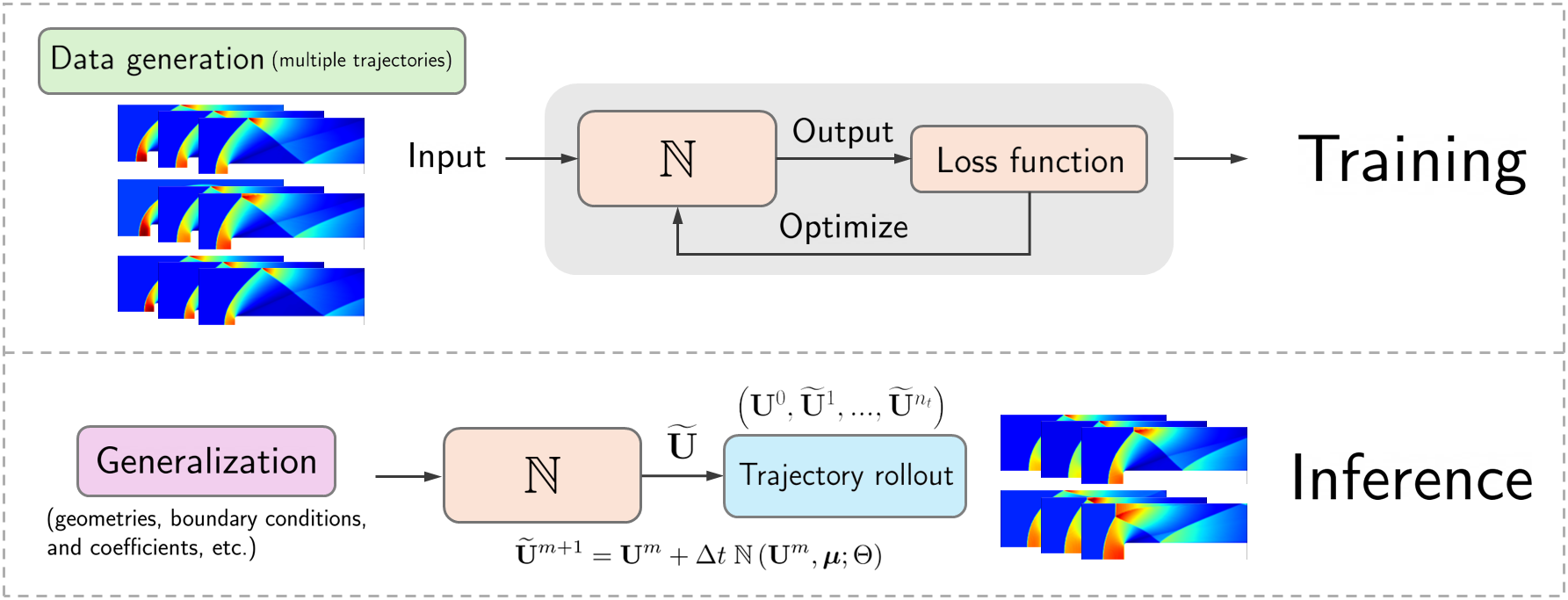}}
	\ \qquad
	\subfloat[Core design components.]{
		\includegraphics[scale=0.36]{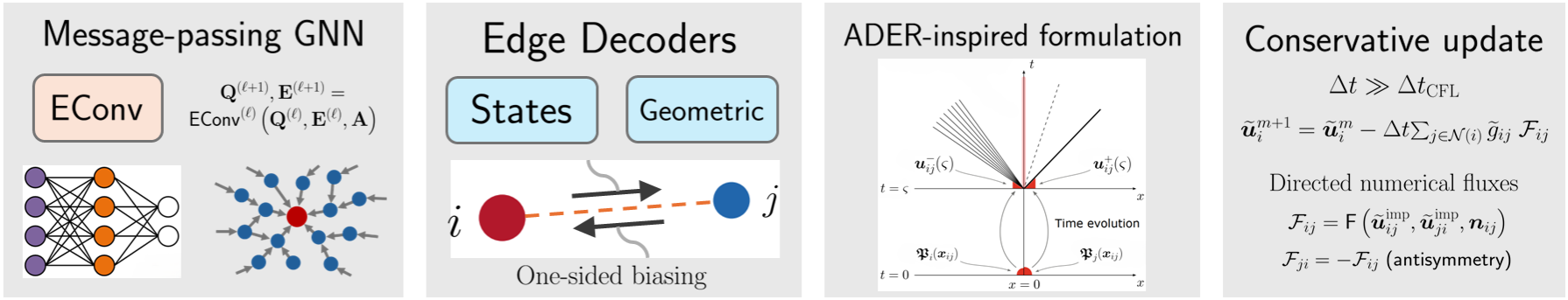}}
	\caption{Overview of the proposed structure-preserving graph neural solver for interpretable modeling of parametric hyperbolic conservation laws. Our framework tightly integrates classical Godunov‑type discretization principles with GNNs, built around three interconnected innovations: 1) Directional message-passing operations emulate wide spatial stencils for high-order nonlinear reconstructions, with one-sided (left/right) biasing representations emerging naturally in the latent space. 2) Rather than a black-box state updater, the network is treated as a learned reconstruction-and-flux operator via edge-wise decoding coupled with a differentiable Riemann solver, enforcing local conservation, upwinding, and entropy-satisfying properties directly in the architecture. 3) Inspired by ADER schemes, the message-passing GNN is further recast as a high-order space-time predictor that captures fine-scale temporal dynamics, yielding one-step implicit-like conservative updates that remain Riemann-solver compatible while operating stably at large time steps.}
	\label{fig:SchematicNeuralSolver}
\end{figure}

Our approach consists of two main stages, as outlined in \textbf{Figure~\ref{fig:SchematicNeuralSolver}}, which provides an overview of the proposed structure‑preserving solver framework. The training stage requires building a dataset of multiple trajectories generated with the high-fidelity solver under varying PDE parameters, which serve as ground-truth labels. Once trained, $\mathbb{N} \left ( \cdot ; \Theta \right )$ is applied autoregressively during inference, starting from an initial state $\mathbf{U}^{0}$ to produce the entire rollout trajectory $\left ( \mathbf{U}^{0}, \widetilde{\mathbf{U}}^{1}, ..., \widetilde{\mathbf{U}}^{n_t} \right )$. The neural solver essentially learns to approximate the nonlinear evolution operator of the underlying PDEs, allowing generalization to new PDE parameters unseen in the training data. Compared with classical numerical methods, this learning-based approach enables rapid trajectory rollout, achieving orders of magnitude simulation speedups. Such efficiency gains make it highly promising for large-scale flow problems and engineering applications such as design, optimization, and real-time prediction.

\section{Graph Neural Networks}

We develop data-driven neural solvers based on GNNs to obtain parametric PDE solutions defined on mesh-based or point-cloud representations of spatial domains. In contrast to grid-based neural architectures, GNNs operate natively on unstructured data, making them well-suited for irregular and complex geometries commonly encountered in computational fluid dynamics. By explicitly encoding spatial connectivity and enabling localized information exchange through neighborhood message passing, GNNs provide a flexible and expressive framework for learning nonlinear dynamics while respecting the underlying geometric structure of the simulation domain \cite{zhao2024review}.

\subsection{Graph Representation}

\begin{figure}[!htb]
	\centering
	\subfloat[Cell-centered]{
		\includegraphics[scale=0.28]{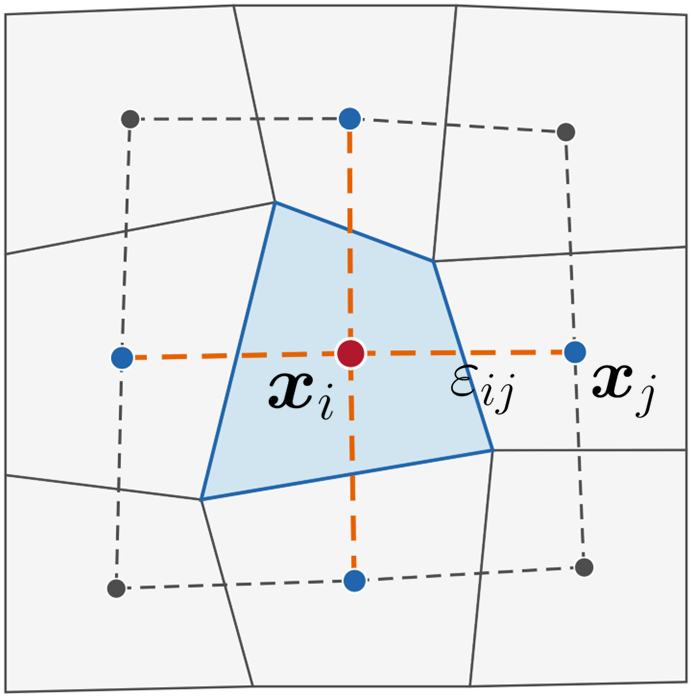}}
	\ \qquad
	\subfloat[Vertex-centered]{
		\includegraphics[scale=0.27]{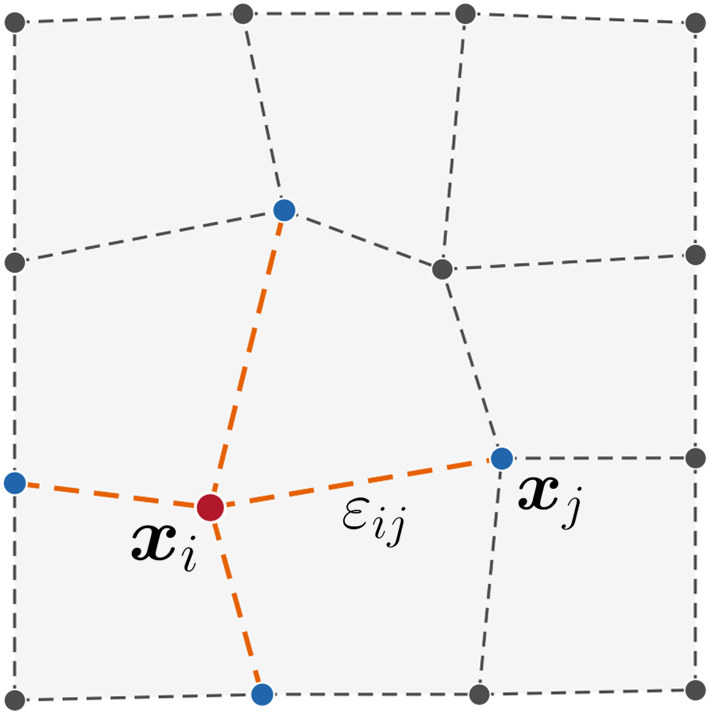}}
	\caption{Schematic of graph representations for a simulation mesh. Dots denote nodes, and dashed lines denote edges (connections between neighboring cells or vertices).}
	\label{fig:GraphRepre}
\end{figure}

A computational mesh can be represented as an undirected graph $\mathcal{G} = \left ( \mathcal{X}, \mathcal{E} \right )$, as shown in \textbf{Figure~\ref{fig:GraphRepre}}. For a cell-centered FV discretization, the node set $\mathcal{X}$ consists of elements $i \in \mathcal{X}$, where each corresponds to a control volume $\Omega_i$. The spatial coordinate of node $i$, denoted by $\bm{x}_i$, is taken as the centroid of $\Omega_i$. The edge set $\mathcal{E}$ contains undirected (unique) edges. Each unique edge $\varepsilon_{ij}$ connects adjacent nodes (cells) $(i,j)$ and is represented in computations by two directed edges $(i \to j)$ and $(j \to i)$. We further denote by $\mathcal{N}(i)$ the set of nodes neighboring node $i$.

\subsection{Generic message passing layer}

The building block of GNNs is the \emph{message passing} mechanism, in which node features are iteratively updated through information exchange with neighbors \cite{gilmer2017neural}. Let $\bm{q}_i$ denote the feature vector associated with node $i$, and let $\bm{e}_{ij}$ represent the feature vector carried by a directed edge $(i \to j)$, encoding pairwise geometric or physical interactions. Within a single message passing layer, node features are updated by aggregating messages from neighboring nodes through a permutation-invariant operation $\bigoplus$ (such as summation or averaging), combined with learnable nonlinear transformations. In general, the update rule can be written as
\begin{equation}
	{\bm{q}}'_i = \phi_\upsilon \left ( \bm{q}_i, \underset{j\in\mathcal{N}(i)}{\bigoplus} \phi_e \left ( \bm{q}_i, \bm{q}_j, \bm{e}_{ij} \right ) \right ),
\end{equation}
where $\phi_e(\cdot)$ computes edge-wise messages from each neighbor $j$ to the target node $i$, and $\phi_\upsilon(\cdot)$ integrates the aggregated messages with the original node features.

We collect the node and edge features into matrices $\mathbf{Q} \in \mathbb{R}^{n \times \hbar_\upsilon}$ and $\mathbf{E} \in \mathbb{R}^{2n_e \times \hbar_{e}}$, respectively, where each row corresponds to a feature vector. Here, $n$ denotes the number of nodes, $n_e$ the number of unique edges, and $\hbar_\upsilon$, $\hbar_{e}$ the feature dimensions. The MPNN layer in compact matrix form that updates both node and edge features for $\ell=0, ..., L-1.$ can then be expressed as
\begin{equation}
	\label{eq:mMPNN}
	\mathbf{Q}^{(\ell+1)}, \mathbf{E}^{(\ell+1)} = \mathsf{MPNN}^{(\ell)} \left ( \mathbf{Q}^{(\ell)}, \mathbf{E}^{(\ell)}, \mathbf{A} \right ),
\end{equation}
where $\mathbf{A} \in \mathbb{R}^{n \times n}$ is the sparse adjacency matrix encoding the graph connectivity.

In this work, we consider a specific realization of Eq.~(\ref{eq:mMPNN}), with its update equations given by
\begin{equation}
	\widehat{\bm{e}}^{(\ell+1)}_{ij} = \phi_e \left ( \bm{q}^{(\ell)}_i, \bm{q}^{(\ell)}_j, \bm{e}^{(\ell)}_{ij} \right ) = \mathsf{MLP}^{(\ell)}_e \left ( [ \, \bm{q}^{(\ell)}_i \, | \, \bm{q}^{(\ell)}_j \, | \, \bm{e}^{(\ell)}_{ij} \, ] \right ),
\end{equation}
\begin{equation}
	\label{eq:nu_MPNN}	
	\widehat{\bm{q}}^{(\ell+1)}_i = \phi_\upsilon \left ( \bm{q}^{(\ell)}_i, \sum_{j\in\mathcal{N}(i)} \! \widehat{\bm{e}}^{(\ell+1)}_{ij} \right ) = \mathsf{MLP}^{(\ell)}_\upsilon \left ( [ \, \bm{q}^{(\ell)}_i \, | \, \sum_{j\in\mathcal{N}(i)} \! \widehat{\bm{e}}^{(\ell+1)}_{ij} \, ] \right ),
\end{equation}
\begin{equation}
	\bm{q}^{(\ell+1)}_i = \widehat{\bm{q}}^{(\ell+1)}_i + \bm{q}^{(\ell)}_i, \quad 
	\bm{e}^{(\ell+1)}_{ij} = \widehat{\bm{e}}^{(\ell+1)}_{ij} + \bm{e}^{(\ell)}_{ij},
\end{equation}	
where $|$ denotes vector concatenation. Residual connections \cite{he2016deep} are employed for both node and edge features to improve training stability and facilitate the construction of deep architectures.

By composing several message passing layers with independent parameters, the effective receptive field of each node progressively expands from immediate neighbors (one-ring connectivity) to distant regions of the graph, enabling the network to capture longer-range dependencies.

\subsection{Edge Convolution Layer}

We further propose an \emph{edge convolutional} message passing layer by establishing an analogy with FV schemes. The update rules take the following form
\begin{equation}
	\label{eq:edgeconv1}
	\widehat{\bm{e}}^{(\ell+1)}_{ij} = \mathsf{MLP}^{(\ell)}_e \left ( [ \, \bm{q}^{(\ell)}_i \, | \, \bm{q}^{(\ell)}_j \, | \, \bm{e}^{(\ell)}_{ij} \, ] \right ),
\end{equation}

\begin{equation}
	\label{eq:nu_EC}	
	\widehat{\bm{q}}^{(\ell+1)}_i = \mathbf{W}^{(\ell)}_\upsilon \, \bm{q}^{(\ell)}_i + \sum_{j\in\mathcal{N}(i)} \! \widehat{\bm{e}}^{(\ell+1)}_{ij},
\end{equation}

\begin{equation}
	\bm{q}^{(\ell+1)}_i = \mathsf{ReLU} \left ( \mathsf{LayerNorm} \left ( \widehat{\bm{q}}^{(\ell+1)}_i \right ) \right ),
\end{equation}	

\begin{equation}
	\label{eq:edgeconv4}
	\bm{e}^{(\ell+1)}_{ij} = \widehat{\bm{e}}^{(\ell+1)}_{ij} + \bm{e}^{(\ell)}_{ij},
\end{equation}	
where $\mathbf{W}_\upsilon$ is a learnable weight matrix. This design discards $\mathsf{MLP}_\upsilon$, thereby reducing the cost of Eq.~(\ref{eq:nu_MPNN}), and can be viewed as a learnable skip connection. The layer is expressed compactly as
\begin{equation}
	\label{eq:DeEConv}
	\mathbf{Q}^{(\ell+1)}, \mathbf{E}^{(\ell+1)} = \mathsf{EConv}^{(\ell)} \left ( \mathbf{Q}^{(\ell)}, \mathbf{E}^{(\ell)}, \mathbf{A} \right ).
\end{equation}

To better guide the network toward learning local nonlinear interactions of hyperbolic PDEs, we further extend the above formulation into an \emph{enriched edge convolution} that emphasizes discrete differential operators on graph edges. Concretely, the edge inputs are enriched by incorporating a difference term between adjacent nodes, yielding a more expressive representation of local gradients. The edge message function becomes
\begin{equation}
	\label{eq:enEdgeFunc}
	\widehat{\bm{e}}^{(\ell+1)}_{ij} = \mathsf{MLP}^{(\ell)}_e \left ( [ \, \bm{q}^{(\ell)}_i \, | \, \bm{q}^{(\ell)}_j \, | \, \bm{e}^{(\ell)}_{ij} \, | \, (\bm{q}^{(\ell)}_j - \bm{q}^{(\ell)}_i) \, ] \right ).
\end{equation}

\subsection{Encode-Process-Decode Architecture} \label{basicEPD}

Encode-Process-Decode (EPD) architectures are commonly employed to build neural surrogate models for predicting the next-step dynamic state $\widetilde{\mathbf{U}}^{m+1}$ of PDEs. For a simulation graph, the input node features consist of the current state variables $\mathbf{U}^{m}$ or $\widetilde{\mathbf{U}}^{m}$ at each timestep alongside geometric and PDE-related information. Edge features encode pairwise relationships between adjacent nodes.

Dedicated node and edge Encoders concatenate the raw inputs and project them into initial latent embeddings $\mathbf{Q}^{(0)} \in \mathbb{R}^{n \times \hbar_\upsilon}$ and $\mathbf{E}^{(0)} \in \mathbb{R}^{2n_e \times \hbar_{e}}$, offering essential physical and geometric representations of flow problems. The Processor module then applies a stack of message passing layers to iteratively refine the latent features. Finally, the Decoder maps the refined node embeddings after the last processing layer to the nodal predictions of the target state $\widetilde{\mathbf{U}}^{m+1}$.

\section{Godunov Graph Neural Solvers}

Data-driven neural surrogates in the form of Eq.~(\ref{eq:disupdfor}), when built from black-box neural network models, lack the capability to preserve the intrinsic structures of the underlying conservation laws. Consequently, they are prone to produce non-physical solutions, generalize poorly, and lose predictive reliability during long-term rollouts. In this work, we pursue a novel, interpretable neural solver design by integrating the rigor of classical Godunov numerical principles (Section~\ref{GodunovFramework}) with the expressive power of GNNs. The resulting framework enforces discrete conservation and key properties such as upwinding at the architectural level (hard constraints).

\subsection{Numerical Fluxes}

Numerical fluxes (Riemann solvers) form the foundation of Godunov-type methods, and are crucial for ensuring upwinding and stability when solving the Euler equations \cite{toro2013riemann}. A suitable numerical flux function should satisfy several key properties:
\begin{itemize}[leftmargin=1em]
	\item \textit{Conservation.}
	The numerical flux must be antisymmetric with respect to the exchange of its left and right arguments
	\begin{equation} 
		\mathsf{F} \left ( \bm{u}^{-}, \bm{u}^{+}, \bm{n} \right ) = -\,\mathsf{F} \left ( \bm{u}^{+}, \bm{u}^{-}, -\bm{n} \right )
	\end{equation}
	so that fluxes across an interface cancel exactly between neighboring cells.
	
	\item \textit{Consistency.}
	Consistent with the physical flux
	\begin{equation} 
		\mathsf{F} \left ( \bm{u}, \bm{u}, \bm{n} \right ) = \mathbf{F}(\bm{u})\cdot \bm{n}
	\end{equation}
	guaranteeing the correct continuous limit in smooth regions of the flow.
	
	\item \textit{Lipschitz continuity.}
	Lipschitz continuous with respect to its arguments.
	
	\item \textit{Rankine–Hugoniot consistency.}
	Enforce the Rankine–Hugoniot (jump) relation across discontinuities, thereby capturing the correct shock propagation and wave speeds.
	
	\item \textit{Entropy satisfaction.}
	Ensure that discontinuities are captured in a physically consistent way, avoiding non-admissible solutions and helping suppress oscillations.
	
\end{itemize}

Common approximate Riemann solvers that satisfy the above properties include the Rusanov \cite{Rusanov1962TheCO}, HLL-family \cite{harten1983upstream,toro1994restoration} and Osher–Solomon \cite{osher1982upwind} fluxes. In this work, we adopt the general central formulation. For a given cell interface $(ij)$, the numerical flux in the normal direction $\bm{n}_{ij}$ is written as
\begin{equation} 
	\mathsf{F} \left ( \bm{u}_{ij}^{-}, \bm{u}_{ij}^{+}, \bm{n}_{ij} \right ) = \frac{1}{2}\left ( \mathbf{F}(\bm{u}_{ij}^{-}) + \mathbf{F}(\bm{u}_{ij}^{+}) \right )\cdot \bm{n}_{ij} - \frac{1}{2} \mathcal{D}_{ij} \, (\bm{u}_{ij}^{+} - \bm{u}_{ij}^{-}),
\end{equation}
where $\mathcal{D}_{ij}$ in the upwinding term is a dissipation matrix depending on the local states and properties of the flux Jacobian $\nabla_{\bm{u}} \mathbf{F}(\bm{u})$. Its specific definition characterizes different Riemann solvers.

A well-known example is the Roe solver \cite{roe1981approximate}, which employs the Roe-averaged matrix $\mathcal{D}_{ij} = \left | \mathcal{A}(\bm{u}_{ij}^{-}, \bm{u}_{ij}^{+}) \right |$. The Roe flux achieves good resolution of contact and shear waves owing to its characteristic decomposition, but is not strictly entropy-satisfying; thus, an entropy fix is typically required. A popular alternative is the Rusanov flux \cite{Rusanov1962TheCO}, which uses a scalar multiple of the identity $\mathcal{D}_{ij} = \mathrm{max}\left\{ \lambda(\bm{u}_{ij}^{-}), \lambda(\bm{u}_{ij}^{+}) \right\} \mathbf{I}$, with $\lambda(\bm{u})$ denoting the maximal absolute eigenvalue of the flux Jacobian. This yields a simple and entropy-stable flux based on an upper bound for propagation speed.

\subsection{Neural Spatial Reconstruction}

\begin{figure}[!htb]
	\centering
	\subfloat[Large spatial stencil for reconstruction]{
		\includegraphics[scale=0.17]{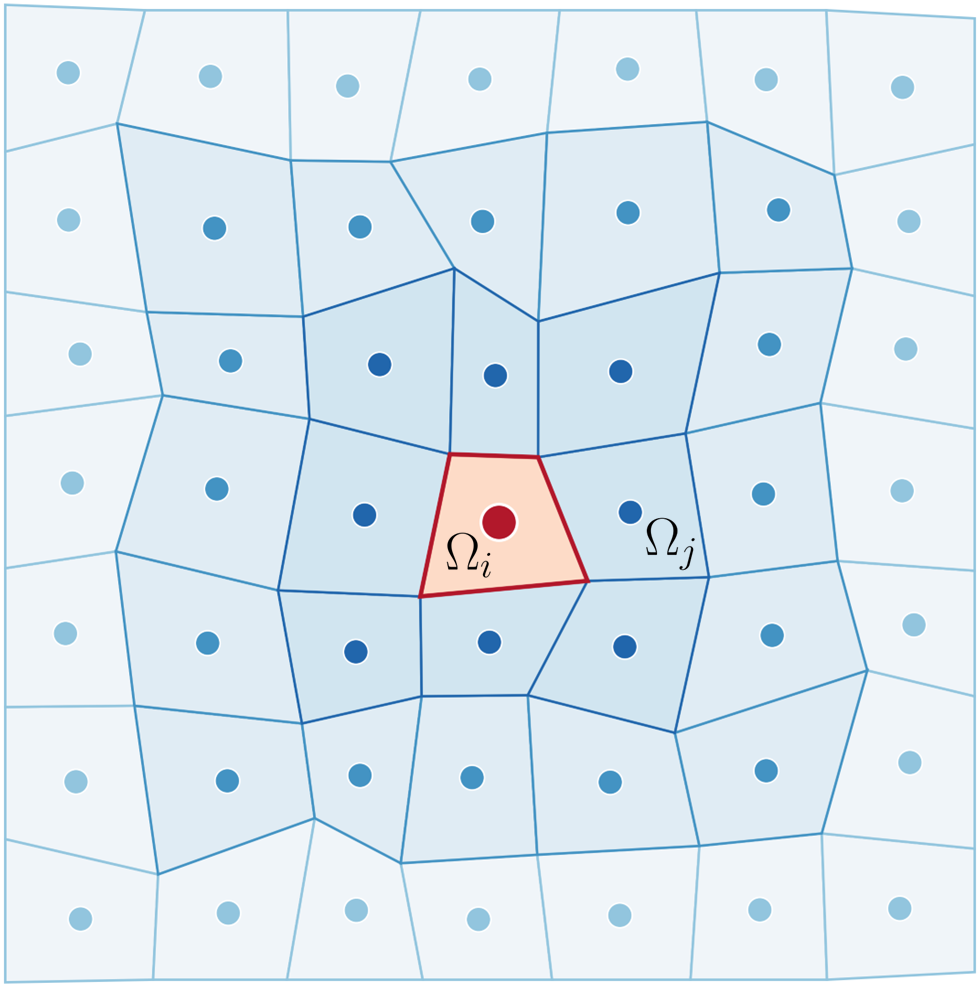}}
	\quad
	\subfloat[Successive message-passing operations]{
		\raisebox{0.06\height}{
			\includegraphics[scale=0.19]{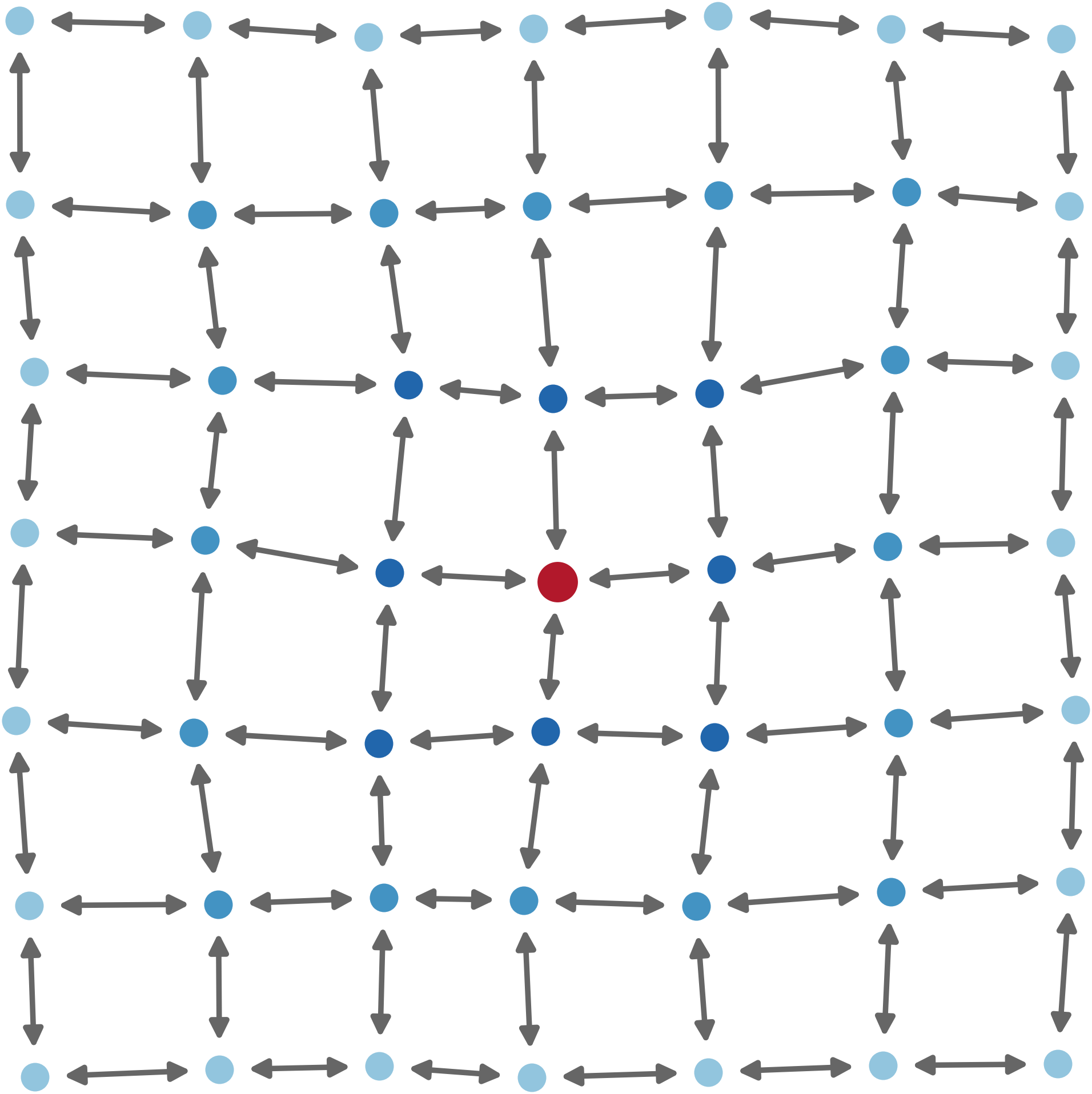}}}
	\qquad
	\subfloat[Directed edges and edge-wise decoding to obtain left/right interface states.]{
		\raisebox{0.26\height}{
			\includegraphics[scale=0.3]{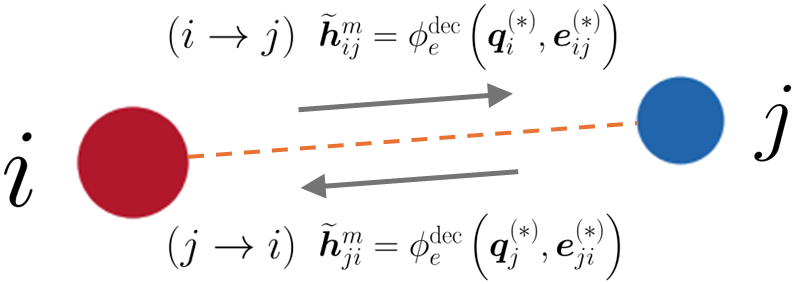}}}
	\caption{Schematics of neural reconstruction process via message-passing GNN as a counterpart to WENO scheme (the target cell is marked with red color).}
	\label{fig:WENOGNN}
\end{figure}

In high-order Godunov-type methods, spatial reconstruction is applied to obtain the left and right states at each cell interface (as described in Section~\ref{HighorderRecons}), ensuring upwinding compatible with Riemann solvers. The nonlinear reconstruction process can be learned through a GNN,  which operates on unstructured meshes and encodes local connectivity, acting as a data-driven counterpart to high-order schemes such as MUSCL or WENO. Through successive geometry-aware message-passing operations, each node aggregates information from its neighbors, effectively emulating the action of large, spatially biased stencils (as illustrated in \textbf{Figure~\ref{fig:WENOGNN}}). Because message propagation along directed graph edges is asymmetric, one-sided (left/right) biasing emerges naturally within the latent embeddings.

Compared to node-wise decoders commonly used in neural surrogates, here we rely on an edge-wise decoder $\phi^{\mathrm{dec}}_e$ to predict the states on both sides of each interface $(i \to j)$. Specifically, given the processed node and edge features $\bm{q}^{(\ast)}$ and $\bm{e}^{(\ast)}$ from the GNN, the decoder outputs component–wise primitive variables
\begin{equation}
	\widetilde{\bm{h}}_{ij}^{m} = \phi^{\mathrm{dec}}_e \left ( \bm{q}^{(\ast)}_i, \bm{e}^{(\ast)}_{ij} \right ),
	\qquad
	\widetilde{\bm{h}}_{ji}^{m} = \phi^{\mathrm{dec}}_e \left ( \bm{q}^{(\ast)}_j, \bm{e}^{(\ast)}_{ji} \right ).
\end{equation}
These quantities serve as neural analogs of $\bm{u}_{ij}^{-}$ and $\bm{u}_{ij}^{+}$ in the classical Godunov formulation. When converted to conserved variables $\widetilde{\bm{u}}_{ij}^{m}$ and $\widetilde{\bm{u}}_{ji}^{m}$, they are then input to a Riemann solver to evaluate the numerical flux.

The GNN essentially learns to produce limited (or non-oscillatory) interface approximations from the known cell averages at the current timestep $m$. The subsequent first-order explicit conservative FV update of the target node (cell) $i$ follows
\begin{equation} 
	\widetilde{\bm{u}}_{i}^{m+1} = \widetilde{\bm{u}}_{i}^{m} - \Delta t \! \sum_{j\in\mathcal{N}(i)} \frac{|s_{ij}|}{|\Omega_i|} \; \mathsf{F} \left ( \widetilde{\bm{u}}_{ij}^{m}, \widetilde{\bm{u}}_{ji}^{m}, \bm{n}_{ij} \right ).
\end{equation}

\section{ADER-inspired Neural Solver}

The GNN-based reconstruction proposed earlier effectively replaces the cumbersome and costly computations of high-order polynomials and nonlinear weights required in WENO. However, each FV update depends only on the solutions at $t^m$. Consequently, a purely learned spatial reconstruction remains bound by the stability (CFL) restrictions of explicit integration, enforcing timestep sizes on the order of $10^{-4} \sim 10^{-5}$. This severely limits training and inference efficiency of the neural solver, diminishing potential speed-ups over classical numerical solvers. To enable conservative and stable updates with large timesteps, we instead treat the GNN as a high-order space-time predictor, and construct a one-step implicit-like neural solver inspired by the Arbitrary high order DERivatives (ADER) methodology.

\subsection{ADER schemes}

ADER schemes \cite{titarev2002ader} are high-order extensions of Godunov methods and rely on two building blocks: (1) spatial reconstruction and (2) solution of generalized Riemann problems (GRPs) at cell interfaces for flux evaluation. The ADER formulation begins by integrating the conservation laws over the space-time control volume $\Omega_i\times[t^m,t^{m+1}]$, leading to the one-step FV update
\begin{equation} 
	\bm{u}_{i}^{m+1} = \bm{u}_{i}^{m} - \Delta t \! \sum_{j\in\mathcal{N}(i)} \frac{|s_{ij}|}{|\Omega_i|} \; \mathcal{F}_{ij},
\end{equation}
where $\mathcal{F}_{ij}$ denotes the time-integral average of fluxes
\begin{equation}
	\mathcal{F}_{ij} = \frac{1}{\Delta t} \int_{t^m}^{t^{m+1}} \! \! \mathbf{F}\Big ( \bm{u}(\bm{x}_{ij}, t) \Big )\cdot \bm{n}_{ij}\, \mathrm{d}t.
\end{equation}

ADER employs a set of polynomials $\left\{ \bm{\mathfrak{P}}_{i}(\bm{x}) \right\}$ of degree $K$, provided by nonlinear reconstruction procedure such as WENO. At each interface, the reconstructed polynomials define the initial conditions for a GRP \cite{toro2002solution,montecinos2012comparison} posed in the local normal coordinate
\begin{equation}
	\begin{aligned}
		& \frac{\partial \bm{u}(\tau, t)}{\partial t} + \frac{\partial \big ( \mathbf{F}(\bm{u})\cdot \bm{n} \big )}{\partial \tau} = 0,  \\
		& \bm{u}(\tau, 0) = 
		\begin{cases}
			\, \bm{\mathfrak{P}}^{-}(\tau), & \tau<0, \\
			\, \bm{\mathfrak{P}}^{+}(\tau), & \tau>0.
		\end{cases}
	\end{aligned}
\end{equation}

One strategy available for solving this GRP is the HEOC solver \cite{castro2008solvers}, which consists of a temporal evolution step and a data interaction step. The evolution step evolves the interface states via Taylor series expansions in time $t=\varsigma$
\begin{equation}
	\left\{
	\begin{aligned}
		& \bm{u}_{ij}^{-}(\varsigma) = \bm{\mathfrak{P}}_{i}(\bm{x}_{ij}) + \sum_{k=1}^{K} \partial^{(k)}_t \bm{\mathfrak{P}}_{i}(\bm{x}_{ij}) \, \frac{\varsigma^{k}}{k!}, \\
		& \bm{u}_{ij}^{+}(\varsigma) = \bm{\mathfrak{P}}_{j}(\bm{x}_{ij}) + \sum_{k=1}^{K} \partial^{(k)}_t \bm{\mathfrak{P}}_{j}(\bm{x}_{ij}) \, \frac{\varsigma^{k}}{k!},
	\end{aligned}
	\right.
\end{equation}
where $\varsigma=t-t^m$ denotes the local time. Through the Cauchy-Kovalevskaya procedure, the high-order time derivatives are replaced by spatial derivatives.

During the subsequent data interaction step, a classical Riemann problem is defined by the left and right evolved states, which serve as inputs to an approximate Riemann solver $\mathsf{F} \left ( \bm{u}_{ij}^{-}(\varsigma), \bm{u}_{ij}^{+}(\varsigma), \bm{n}_{ij} \right )$. Finally, the time-averaged numerical flux $\mathcal{F}_{ij}$ is obtained by approximating the time-integral via a suitable quadrature rule, achieving high-order accuracy in both space and time.

\subsection{Space-time Formulation}

Within the Godunov framework, solutions are typically advanced through explicit FV updates constrained by the CFL condition. To circumvent this and achieve accurate predictions on temporally sparse trajectories with large timestep sizes $\Delta t \gg \Delta t_{\mathrm{CFL}}$, we reinterpret the message-passing GNN as a space-time predictor rather than a purely spatial reconstructor.

The central idea is to let the GNN emulate the ADER mechanism: instead of generating instantaneous reconstructions at $t^m$, the network learns to approximate high-order polynomials that encode temporal evolution over each time interval. Accordingly, the edge-wise decoder $\phi^{\mathrm{dec}}_e$, analogous to the ADER-HEOC solver, is extended to output left/right interface states
\begin{equation}
	\label{eq:starfeatures}
	\widetilde{\bm{h}}_{ij}^{\mathrm{imp}} = \phi^{\mathrm{dec}}_e \left ( \bm{q}^{(\ast)}_i, \bm{e}^{(\ast)}_{ij} \right ),
	\qquad
	\widetilde{\bm{h}}_{ji}^{\mathrm{imp}} = \phi^{\mathrm{dec}}_e \left ( \bm{q}^{(\ast)}_j, \bm{e}^{(\ast)}_{ji} \right ),
\end{equation}
which embed both space and time high-order information.

During flux evaluation, the predicted states are transformed into conserved variables and supplied to a numerical flux function. The resulting fluxes correspond to integral averages rather than quantities at $t^m$. The discrete update then follows as
\begin{equation}
	\label{eq:ADERdiscreteUpd}
	\widetilde{\bm{u}}_{i}^{m+1} = \widetilde{\bm{u}}_{i}^{m} - \Delta t \! \sum_{j\in\mathcal{N}(i)} \widetilde{g}_{ij} \; \mathcal{F}_{ij}, \qquad \mathcal{F}_{ij} = \mathsf{F} \left ( \widetilde{\bm{u}}_{ij}^{\mathrm{imp}}, \widetilde{\bm{u}}_{ji}^{\mathrm{imp}}, \bm{n}_{ij} \right ),
\end{equation}
yielding a one-step implicit-like neural solver that remains conservative and stable even for time steps far exceeding the explicit CFL bound, as fine-scale dynamics are absorbed into the space-time latent representations. The resulting formulation retains the strengths of the Godunov framework—conservation, upwinding, and Riemann-solver compatibility. The computation of the geometric weights $\widetilde{g}_{ij}$ is described in Section~\ref{NetDecoders}.

In summary, the proposed ADER-inspired neural solver unifies spatial reconstruction and temporal evolution into a single end-to-end learnable operator, transforming the GNN into a one-step space-time predictor capable of capturing complex wave interactions over large time intervals.

Trajectory data for training the solver are generated using high-order numerical solvers. Temporal coarsening is then performed by sampling the reference solutions at larger time intervals, creating new trajectories with fewer timesteps. Each coarse interval embodies the cumulative effect of high-order space–time discretizations, allowing the GNN to learn the integral behavior of the underlying solver.

\section{Network Architecture Details} \label{NetArchi}

\begin{figure}[!htb]
	\centering
	\includegraphics[scale=0.35]{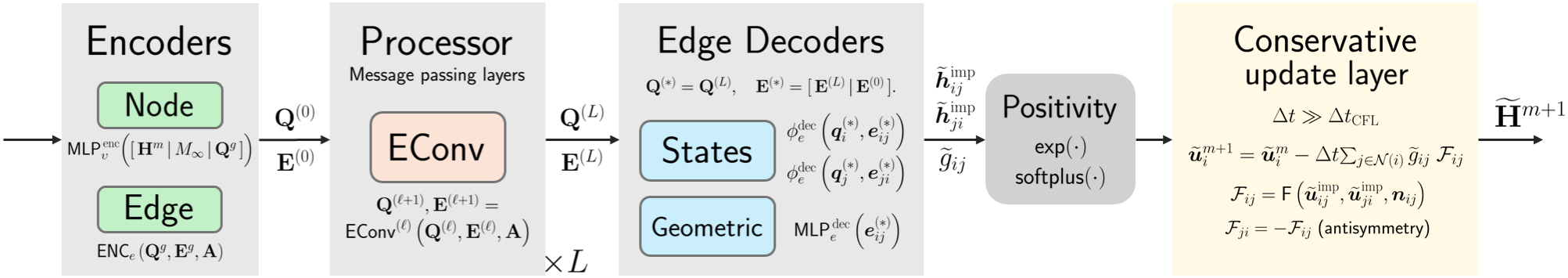}
	\caption{Schematic of the forward pass through the proposed EPD architecture (CPGNet).}
	\label{fig:architecture}
\end{figure}

Our conservation-preserving Godunov-type (CPG) neural solver relies on an EPD architecture (as illustrated in \textbf{Figure~\ref{fig:architecture}}) tailored to point-cloud graphs. All components utilize two-hidden-layer MLPs with $\hbar = 128$ and ReLU nonlinearities (unless otherwise specified). Layer normalization \cite{ba2016layer} is applied to all MLP outputs except the decoders. The architecture concludes with a conservative update layer.

Input features and target states are standardized to zero mean and unit variance using dataset statistics for training stability. Within the conservative framework, both predictions and ground-truth targets are passed through a shared output normalizer before loss computation to ensure consistent scaling.

\subsection{Encoder}

To initialize latent representations that incorporate both the physical flow states and geometric information, we employ dedicated node and edge encoders. For a point-cloud graph, each node implicitly corresponds to a control volume, with input features comprising the current primitive state $\mathbf{H}^{m}$ and geometric context. The concatenated input is processed by an MLP to produce initial node embeddings as
\begin{equation}
	\bm{q}^{(0)}_i = \mathsf{MLP}^{\,\mathrm{enc}}_\upsilon \Big ( [ \, \bm{h}_i^m \, | \, M_\infty \, | \, \bm{q}^{g}_i \, ] \Big ),
\end{equation}
where $M_\infty$ denotes the free-stream Mach number treated as a global condition appended to all nodes. The geometric vector $\bm{q}^{g}_i = [ \, \bm{x}_i \, | \, \bm{\eta}_i \, ]$ includes the node coordinates $\bm{x}_i$ and a one-hot encoding $\bm{\eta}_i$ indicating the node type (e.g., interior, wall, inflow, outflow).

The edge Encoder integrates pairwise relations between nodes. Each directed edge is assigned a geometric feature vector
\begin{equation}
	\bm{e}^{g}_{ij} = [ \, \Delta\bm{x}_{ij} \, | \, d_{ij} \, | \, \bm{n}_{ij} \, ], \qquad \bm{n}_{ij} = \Delta\bm{x}_{ij}/d_{ij},
\end{equation}
where $\Delta\bm{x}_{ij} = \bm{x}_j - \bm{x}_i$ is the relative position vector and $d_{ij} = |\Delta\bm{x}_{ij}|$ is the Euclidean distance. This relative positional encoding preserves spatial equivariance, which is crucial for capturing directional flow patterns.

The initial edge feature matrix is then obtained as
\begin{equation}
	\mathbf{E}^{(0)} = \mathsf{ENC}_e \left ( \mathbf{Q}^{g}, \mathbf{E}^{g}, \mathbf{A} \right ),
\end{equation}
where the encoder $\mathsf{ENC}_e$ consists of three stacked message passing layers, each with the hidden dimension of $\hbar^{\mathrm{enc}}_{e} = 32$.

Together, these encoders transform raw simulation data into rich latent representations that serve as the input to the subsequent GNN processor. The design allows the network to operate directly on point clouds while effectively handling varying geometries and boundary configurations.

\subsection{Processor}

The Processor serves as the core module of the network. It is built from a stack of $L=12$ $\mathsf{EConv}$ layers, each incorporating a directional difference term in the message function (Eq.~\ref{eq:enEdgeFunc}). The resulting node and edge features are denoted by $\mathbf{Q}^{(L)}$ and $\mathbf{E}^{(L)}$, respectively, with the processed embeddings defined as
\begin{equation}
	\mathbf{Q}^{(\ast)} = \mathbf{Q}^{(L)}, \qquad \mathbf{E}^{(\ast)} = [ \, \mathbf{E}^{(L)} \, | \, \mathbf{E}^{(0)} \, ].
\end{equation}
Here, the final and initial edge features are fused to augment the learned space-time couplings with geometric priors for the downstream decoders.

To emulate upwind-biased reconstructions of Godunov-type methods, $\mathbf{Q}^{(\ast)}$ and $\mathbf{E}^{(\ast)}$ are concatenated to form directed edge embeddings, which are then supplied to the edge-wise decoder $\phi^{\mathrm{dec}}_e$, as in Eq.~(\ref{eq:starfeatures}).

\subsection{Decoder} \label{NetDecoders}

There are two decoders. The first, an edge-wise decoder $\phi^{\mathrm{dec}}_e$, predicts implicit left/right interface states $\widetilde{\bm{h}}_{ij}^{\mathrm{imp}}$ and $\widetilde{\bm{h}}_{ji}^{\mathrm{imp}}$, representing space-time high-order reconstructions. Separate lightweight MLPs with hidden dimension $\hbar^{\mathrm{dec}}_{e} = 32$ are used for each primitive variable. Exponential maps $\mathsf{exp}(\cdot)$ enforce positivity of density ($\rho>0$) and pressure ($p>0$), and numerical fluxes are then evaluated through a differentiable Riemann solver—Rusanov in our studies.

The second decoder, also edge-wise, estimates positive geometric weights using a Softplus mapping
\begin{equation}
	\widetilde{g}_{ij} = \mathsf{softplus}\left ( \mathsf{MLP}^{\,\mathrm{dec}}_e \left ( \bm{e}^{(\ast)}_{ij} \right ) \right ) + \epsilon,
\end{equation}
which learns to approximate the ratio ${|s_{ij}|}/{|\Omega_i|}$. A constant $\epsilon=10^{-6}$ is added for numerical robustness. The learned geometric weights enable the network to operate directly on irregular point-cloud simulation data without requiring an explicit FV mesh—only graph connectivity and nodal coordinates as inputs. This allows the neural solver to preserve the conservative structure while remaining agnostic to the underlying discretization. Consequently, the solver gains flexibility across different geometries and resolutions and achieves improved efficiency during both training and inference.

\subsection{Conservative Update Layer}

This layer employs a message aggregation operator and serves as the final stage of the neural solver, responsible for advancing the conserved variables. Directed numerical fluxes $\mathcal{F}_{ij}$ are first evaluated along all $n_e$ unique edges, followed by obtaining undirected pairs $\mathcal{F}_{ji}=-\mathcal{F}_{ij}$ (antisymmetry), ensuring that flux contributions across a shared interface cancel exactly between the two connected nodes. The aggregation step (Eq.~\ref{eq:ADERdiscreteUpd}) is implemented as a scatter-add of weighted fluxes (edge messages) over incoming/outgoing directed edges to update each node's state, thereby guaranteeing local conservation by construction.

\section{Training Strategies}

\subsection{One-step Training}

Under the one-step training setting, the learning task is formulated as a sequence of immediate temporal mappings. Specifically, the training dataset consists of input–output pairs at each time level $t^m$, denoted as $\left ( \mathbf{H}_{\wp}^{m} \rightarrow \mathbf{H}_{\wp}^{m+1} \right)$, with $\wp=1,\ldots,n_r$ as the index of the rollout trajectories produced by a high-fidelity numerical solver.

Training is supervised by a mean squared error (MSE) objective, which measures the discrepancy between the predicted primitive variables $\widetilde{\mathbf{H}}_{\wp}^{m+1}$ and their corresponding ground truth labels $\mathbf{H}_{\wp}^{m+1}$. The loss function is defined as
\begin{equation}
	\mathcal{L}_{\mathrm{onestep}}(\Theta) = \frac{1}{n_r} \frac{1}{n_t} \sum_{\wp=1}^{n_r} \sum_{m=0}^{n_t-1} \left \| \widetilde{\mathbf{H}}_{\wp}^{m+1} - \mathbf{H}_{\wp}^{m+1} \right \|_2^2.
\end{equation}
Here, $n_t$ denotes the number of timesteps (temporal snapshots). The network parameters $\Theta$ are optimized via backpropagation using mini-batch stochastic gradient descent. Note that $\widetilde{\mathbf{H}}_{\wp}^{m+1}$ is converted from the updated conserved values
\begin{equation} 
	\widetilde{\mathbf{U}}_{\wp}^{m+1} = \mathbf{U}_{\wp}^{m} + \Delta t \: \mathbb{N} \left ( \mathbf{U}_{\wp}^{m}, \bm{\mu}_{\wp}; \Theta \right ).
\end{equation}

Neural solvers trained solely with the one-step objective are prone to error accumulation during long-term autoregressive inference: small prediction errors can amplify over time, leading to substantial deviations from the true solution trajectory \cite{sanchez2020learning}. To alleviate the mismatch between training and inference input distributions, we inject additive Gaussian noise $\mathfrak{N} \left ( 0, \sigma \right )$ with zero mean and fixed variance into the input states $\mathbf{H}_{\wp}^{m}$ during training. This perturbation strategy exposes the network to mildly corrupted inputs, improving robustness when the solver is driven by its own predictions during inference.

\subsection{Multistep Strategy}

To address the shortcomings inherent to one-step supervision, we employ a multistep training strategy \cite{long2018pde,lam2023learning}. This approach explicitly accounts for temporal error accumulation by unrolling the neural solver over multiple future time levels within each training iteration. A sliding window of length $n_w$ is used to expose the model to longer temporal contexts, enabling it to learn both short-term evolution and longer-horizon dynamics. Concretely, training samples consist of contiguous state sequences $(\mathbf{H}_{\wp}^{m}, \mathbf{H}_{\wp}^{m+1}, \ldots, \mathbf{H}_{\wp}^{m+n_w})$. The corresponding multistep objective aggregates prediction errors across all unrolled steps and is defined as
\begin{equation}
	\mathcal{L}_{\mathrm{multistep}}(\Theta) = \frac{1}{n_r} \frac{1}{n_t-n_w+1} \sum_{\wp=1}^{n_r} \sum_{m=0}^{n_t-n_w} \sum_{k=1}^{n_w} \left \| \widetilde{\mathbf{H}}_{\wp}^{m+k} - \mathbf{H}_{\wp}^{m+k} \right \|_2^2,
\end{equation}
where $\widetilde{\mathbf{H}}_{\wp}^{m+k}$ denotes the primitive values at the $k$-th step of the rollout window initiated at time $t^m$. Within each window, predictions are produced recursively using the solver's own outputs
\begin{equation} 
	\widetilde{\mathbf{U}}_{\wp}^{m+k} = \widetilde{\mathbf{U}}_{\wp}^{m+k-1} + \Delta t \; \mathbb{N} \left( \widetilde{\mathbf{U}}_{\wp}^{m+k-1}, \bm{\mu}_{\wp}; \Theta \right), \quad k=1,\ldots,n_w, \quad \widetilde{\mathbf{U}}_{\wp}^{m} = \mathbf{U}_{\wp}^{m}.
\end{equation}
Stochastic perturbations are applied exclusively to the initial input of each rollout window, with noise amplitudes that are an order of magnitude smaller than those employed during one-step training. Since subsequent states are generated autoregressively from the model’s own predictions, the multistep formulation exposes the network to the distributional shifts and cumulative errors encountered during inference.

\subsection{Two-stage Curriculum}

Directly optimizing the multistep loss from random initialization is often challenging, as the network must simultaneously capture accurate local dynamics and maintain stability over extended temporal horizons \cite{brandstetter2022message}. To improve optimization robustness, we adopt a two-stage curriculum training strategy. In the first stage, the model is pretrained using the one-step loss, providing a stable initialization that emphasizes short-term accuracy. In the second stage, the pretrained network is fine-tuned using the multistep objective over a limited number of epochs, allowing the model to explicitly reduce accumulated rollout errors without incurring excessive optimization difficulty.

GNN models are trained on an NVIDIA L20 GPU using the Adam optimizer \cite{kingma2014adam}. The one-step (teacher-forcing) stage in the two-stage curriculum runs for 15 epochs with a learning rate of $10^{-4}$. Then the multistep stage runs 5 epochs with the learning rate reduced to $10^{-5}$. In both stages, the temporal context window size is fixed at $n_w=3$. Our implementation relies on PyTorch \cite{paszke2019pytorch} and the PyTorch Geometric library \cite{fey2019fast}.

\section{Benchmark Datasets}

We construct high-fidelity datasets targeting four canonical benchmarks of supersonic Euler flows that are challenging even for state-of-the-art numerical methods. All simulations are performed through the Trixi.jl framework \cite{ranocha2021adaptive,schlottke2021purely} on unstructured quadrilateral meshes generated by Gmsh \cite{geuzaine2009gmsh}. The Euler equations with $\gamma=1.4$ are discretized using the DGSEM \cite{kopriva2010quadrature}. Polynomials of degree $K=3$ (fourth-order spatial accuracy) with a Lobatto–Legendre nodal basis are used on all elements, and the Rusanov numerical flux is chosen for surface flux evaluations. An entropy-stable shock-capturing strategy is enabled based on the Hennemann–Gassner indicator \cite{hennemann2021provably}. Time integration advances the semi-discrete system by a third-order adaptive strong-stability-preserving Runge-Kutta (explicit SSPRK43) scheme \cite{kraaijevanger1991contractivity,ranocha2022optimized}, with the Zhang-Shu positivity-preserving limiter \cite{zhang2011maximum} applied at every RK stage. 

Free-stream states are defined by fixed density $\rho_\infty = 1.4$, pressure $p_\infty = 1.0$, and randomly sampled Mach number $M_\infty$. The velocity components are computed via
\begin{equation} 
	v_\infty = M_\infty \sqrt{\gamma p_\infty / \rho_\infty},
	\qquad (v_1, v_2) = (v_\infty, 0.0),
\end{equation}
and these values specify both the initial and Dirichlet inflow conditions.

This configuration enables the systematic construction of parametric datasets across diverse geometries and flow regimes, covering a broad spectrum of unsteady, strongly nonlinear high-speed phenomena. Each dataset comprises 300 training and 20 test trajectories generated with adaptive timestep control on the order of $10^{-4} \sim 10^{-5}$ to satisfy stability and accuracy requirements. High-resolution DGSEM solutions are temporally coarsened by sampling at large intervals ($\Delta t \gg \Delta t_{\mathrm{CFL}}$), yielding a sequence length of $n_t = 80$ per trajectory. Such design provides consistent simulation data for assessing neural solvers under challenging conditions. Key dataset statistics are given in Table~\ref{tab:datasets}.

\begin{table}[!htb]
	\centering
	\caption{Dataset Statistics. Notation: $T$—total simulation time; $\Delta t_{\textrm{neural}}$—coarsened timestep; $n_{\textrm{avg}}$—average node count.}
	\label{tab:datasets}
	\begin{tabular}{|c|c|c|c|c|}
		\hline
		Dataset             & $T$                   & $\Delta t_{\textrm{neural}}$    & $n_{\textrm{avg}}$ & noise scale $\sigma$ $(\rho, v_1, v_2, p)$ \\ \hline
		Supersonic Bump     & 2.0                   & 0.025          & 19k                & (2e-2, 2e-2, 1e-2, 2e-2)           \\ \hline
		Forward Step & 2.0                   & 0.025          & 23k                & (5e-2, 2e-2, 1e-2, 10e-2)           \\ \hline
		Shock Diffraction   & 0.7                   & 0.00875        & 22k                & (4e-2, 2e-2, 2e-2, 8e-2)           \\ \hline
		Supersonic Cylinder & 1.6                   & 0.02           & 24k                & (2e-2, 1e-2, 1e-2, 3e-2)           \\ \hline
	\end{tabular}
\end{table}

\subsection{Supersonic Bump}

The dataset involves supersonic flows through a rectangular channel with a circular‑arc bump on the lower wall \cite{moukalled2001high}. The computational domain is the rectangle $[0,3] \times [0,1]$; the bump is parameterized by its chord length, maximum height, and starting position, generating variation of channel geometries with diverse shock structures and boundary interactions. The top and bottom boundaries are treated as inviscid slip walls (zero normal velocity), the left boundary prescribes an inlet with a free‑stream Mach number sampled randomly from $M_\infty \in [1.5, 2.5]$, and the right side is a non-reflecting outflow boundary, implemented by returning the physical flux computed from interior states. All simulations evolve over the time interval $t \in [0,2]$. A lower‑resolution counterpart of the dataset is additionally generated using a second-order DGSEM discretization ($K=1$), containing only 200 training trajectories.

\begin{figure}[!htb]
	\centering
	\includegraphics[scale=0.36]{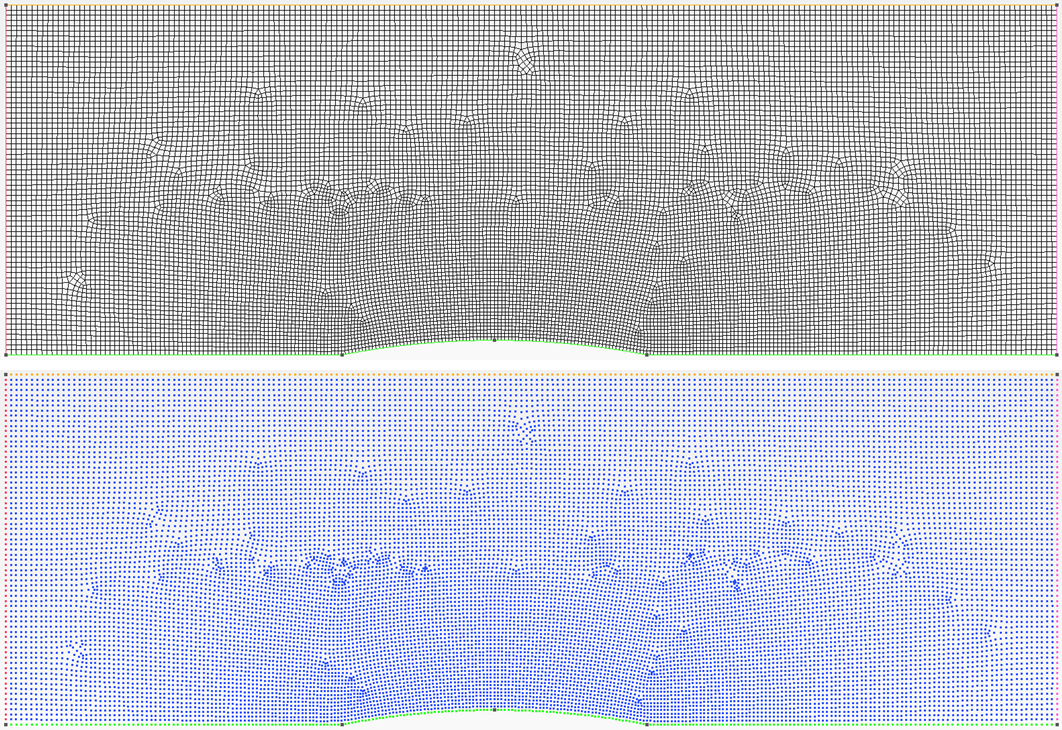}
	\caption{Example DGSEM simulation mesh (top) and the associated point cloud (bottom) for the Supersonic Bump benchmark.}
	\label{fig:MeshPointCloud}
\end{figure}

\textbf{Figure~\ref{fig:MeshPointCloud}} illustrates an example DGSEM simulation mesh and the corresponding point cloud. The point cloud, along with its node connectivity, defines the graph representation that serves as direct input to the proposed graph neural solver (see Section~\ref{NetArchi} for details). This formulation allows the solver to operate purely on point clouds, eliminating the need for an explicit FV mesh during inference.

\subsection{Forward-facing Step}

Supersonic flow past a forward-facing step is a well-known benchmark for assessing high-resolution methods \cite{woodward1984numerical,cockburn1998runge}. While the standard configuration considers a Mach 3 flow in a wind tunnel with a fixed step geometry, we generalize it into a parametric dataset by randomly varying both geometric and flow parameters. The domain is the rectangle $[0,3]\times[0,1]$, in which a sharp step on the lower wall triggers complex phenomena including shock reflections, expansion fans, and flow separation. The step is defined by its streamwise location and height. Boundary conditions are set as: the inflow with $M_\infty\in[2.5,3.5]$ on the left, outflow on the right, and slip-wall conditions along the top, bottom, and step surfaces. All simulations are advanced over $t \in [0,2]$.

\subsection{Shock Diffraction}

Shock diffraction at a sharp backward-facing corner is another popular benchmark for the Euler equations \cite{cockburn1998runge}, known for producing negative density and/or pressure when numerical schemes lack positivity-preserving limiters. The domain is the square $[0,3]\times[0,3]$ containing a rectangular obstacle in the lower-left region, parameterized by its streamwise length and height. A planar shock initially located at $x = 0.3$ moves into undisturbed air. The pre-shock state ($x > 0.3$) is $\bm{h}^{0}_{\mathrm{pre}}=(1.4, 0.0, 0.0, 1.0)^{\top}$, while the post-shock state ($x \leq 0.3$) is obtained via the Rankine–Hugoniot relations
\begin{equation}
	\begin{aligned}
		& \frac{\rho_{\mathrm{po}}}{\rho_{\mathrm{pre}}}=\frac{(\gamma+1)M_\infty^2}{(\gamma-1)M_\infty^2+2},  \\
		& \frac{p_{\mathrm{po}}}{p_{\mathrm{pre}}}=1+\frac{2\gamma}{\gamma+1}(M_\infty^2-1),  \\
		& v_{\mathrm{po}} = M_\infty \sqrt{\gamma \frac{p_{\mathrm{pre}}}{\rho_{\mathrm{pre}}}} \left(1 - \frac{\rho_{\mathrm{pre}}}{\rho_{\mathrm{po}}}\right).
	\end{aligned}
\end{equation}
The initial condition is therefore defined piecewise for each case according to this shock profile. The free-stream Mach number is randomly sampled from $M_\infty\in[2.5,3.5]$.

Boundary conditions include inflow on the left, outflow on the bottom and right, and slip-wall conditions along the top boundary and obstacle surfaces. Note that this differs from the standard setup where the top is also treated as an outflow. The simulation time interval is set to $t \in [0, 0.7]$.

\subsection{Supersonic Cylinder}

The dataset is constructed from a family of supersonic flows past a cylinder \cite{nazarov2017numerical,guermond2018second} embedded in a tunnel of size $[0,4]\times[0,2]$. The obstacle is a closed ellipse whose geometry is parameterized by its center location and semi-axis lengths. Boundary conditions include inflow on the left, outflow on the right, and slip-wall conditions along the top, bottom, and elliptical surfaces. The free-stream Mach number is randomly sampled per case from $M_\infty\in[1.4,2.2]$. This configuration produces a parametric dataset spanning a wide range of shock structures, flow deflections, and wake patterns arising from variations in the Mach number and obstacle geometry. All simulations evolve over $t\in[0,1.6]$.

\section{Results and Evaluation}

We systematically evaluate the accuracy, stability, and generalization performance of our graph neural solver constructed based on the Conservation-preserving Godunov-type network (CPGNet) architecture (see Section~\ref{NetArchi} for architectural details), by comparing it with three strong baselines: (1) \textbf{GINO}: geometry-informed neural operator \cite{li2023geometry}; (2) \textbf{GNOT}: general neural operator transformer \cite{hao2023gnot}; (3) \textbf{MGN} (MeshGraphNet) \cite{pfaff2020learning} configured with 12 MPNN layers. All baselines follow the default settings of their open-source implementations, with model capacities (parameter counts and hidden dimensions) adjusted to match that of CPGNet where applicable. Performance is evaluated both quantitatively, using error metrics computed against high‑fidelity numerical simulations, and qualitatively, by inspecting the solution fields.

\subsection{Quantitative results}

To quantitatively assess rollout accuracy, we measure the cumulative root mean squared error (RMSE) over a full trajectory. Averaging across all timesteps and trajectories in a test set yields the all-step RMSE
\begin{equation}
	\text{RMSE}_{\text{all}} = \frac{1}{n_r} \sum_{\wp=1}^{n_r} \sqrt{\frac{1}{n_t} \sum_{m=0}^{n_t-1} \left \| \widetilde{\mathbf{H}}_{\wp}^{m+1} - \mathbf{H}_{\wp}^{m+1} \right \|_2^2},
\end{equation}
where $\widetilde{\mathbf{H}}_{\wp}^{m+1}$ denotes the autoregressively predicted state. This metric reflects error growth of rollouts during inference, and is reported separately for each primitive variable.

\begin{table}[!htb]
	\centering
	\caption{Rollout errors $\text{RMSE}_{\text{all}}$ comparisons of different methods under the one-step training. Datasets are generated with the second-order DGSEM in (a) and with the fourth-order DGSEM in (b-e). The best results are in \textbf{bold}, the second‑best are \underline{underlined}, and the third-best are $\overline{overlined}$. Across every dataset and test case, the proposed CPGNet consistently outperforms all baseline architectures, achieving the lowest rollout errors. Within the CPGNet processor module, $\mathsf{EConv}$ delivers higher accuracy relative to $\mathsf{GT}$, which in turn surpasses $\mathsf{GAT}$.}
	\label{tab:RMSE_1}
    
	\subfloat[Supersonic Bump (second-order DGSEM)]{
		\centering
		\begin{tabular}{|c|cccccc|}
			\hline
			\multirow{2}{*}{Var.} & \multirow{2}{*}{$\textrm{GINO}$} & \multirow{2}{*}{$\textrm{GNOT}$} &  \multirow{2}{*}{$\textrm{MGN}$} & \multicolumn{3}{c|}{$\textrm{CPGNet}$}                                     \\ \cline{5-7}
			&            &        &                     & \multicolumn{1}{c}{$\mathsf{GAT}$}   & \multicolumn{1}{c}{$\mathsf{GT}$}    & $\mathsf{EConv}$ \\ \hline
			$\rho$       & 0.034  & 0.043  & 0.036      & \multicolumn{1}{c}{$\overline{0.012}$} & \multicolumn{1}{c}{$\underline{0.010}$} & $\mathbf{0.008}$ \\ 
			$v_1$        & 0.015  & 0.020  & 0.013      & \multicolumn{1}{c}{$\overline{0.005}$} & \multicolumn{1}{c}{\underline{0.004}} & $\mathbf{0.003}$ \\ 
			$v_2$        & 0.028  & 0.031  & 0.028      & \multicolumn{1}{c}{$\overline{0.008}$} & \multicolumn{1}{c}{\underline{0.007}} & $\mathbf{0.005}$ \\ 
			$p$          & 0.034  & 0.041  & 0.035      & \multicolumn{1}{c}{$\overline{0.015}$} & \multicolumn{1}{c}{\underline{0.013}} & $\mathbf{0.009}$ \\ \hline
		\end{tabular}
	}
	\quad
	\subfloat[Supersonic Bump]{
		\centering
		\begin{tabular}{|c|cccccc|}
			\hline
			\multirow{2}{*}{Var.} & \multirow{2}{*}{$\textrm{GINO}$} & \multirow{2}{*}{$\textrm{GNOT}$} & \multirow{2}{*}{$\textrm{MGN}$} & \multicolumn{3}{c|}{$\textrm{CPGNet}$}                                     \\ \cline{5-7} 
			&            &        &                      & \multicolumn{1}{c}{$\mathsf{GAT}$}   & \multicolumn{1}{c}{$\mathsf{GT}$}    & $\mathsf{EConv}$ \\ \hline
			$\rho$       & 0.062  & 0.064  & 0.056       & \multicolumn{1}{c}{$\overline{0.021}$} & \multicolumn{1}{c}{\underline{0.018}} & $\mathbf{0.015}$ \\ 
			$v_1$        & 0.042  & 0.045  & 0.035       & \multicolumn{1}{c}{$\overline{0.009}$} & \multicolumn{1}{c}{\underline{0.008}} & $\mathbf{0.007}$ \\ 
			$v_2$        & 0.034  & 0.035  & 0.029       & \multicolumn{1}{c}{$\overline{0.013}$} & \multicolumn{1}{c}{\underline{0.011}} & $\mathbf{0.009}$ \\ 
			$p$          & 0.067  & 0.071  & 0.059       & \multicolumn{1}{c}{$\overline{0.022}$} & \multicolumn{1}{c}{\underline{0.020}} & $\mathbf{0.017}$ \\ \hline
		\end{tabular}
	}
	\\
	\subfloat[Forward Step]{
		\centering
		\begin{tabular}{|c|cccccc|}
			\hline
			\multirow{2}{*}{Var.} & \multirow{2}{*}{$\textrm{GINO}$} & \multirow{2}{*}{$\textrm{GNOT}$} & \multirow{2}{*}{$\textrm{MGN}$} & \multicolumn{3}{c|}{$\textrm{CPGNet}$}                                     \\ \cline{5-7} 
			&            &        &                      & \multicolumn{1}{c}{$\mathsf{GAT}$}   & \multicolumn{1}{c}{$\mathsf{GT}$}    & $\mathsf{EConv}$ \\ \hline
			$\rho$       & 0.473  & 0.421  & 0.337       & \multicolumn{1}{c}{$\overline{0.188}$} & \multicolumn{1}{c}{\underline{0.178}} & $\mathbf{0.134}$ \\ 
			$v_1$        & 0.248  & 0.202  & 0.149       & \multicolumn{1}{c}{$\overline{0.083}$} & \multicolumn{1}{c}{\underline{0.077}} & $\mathbf{0.058}$ \\ 
			$v_2$        & 0.192  & 0.168  & 0.135       & \multicolumn{1}{c}{$\overline{0.085}$} & \multicolumn{1}{c}{\underline{0.078}} & $\mathbf{0.062}$ \\ 
			$p$          & 0.817  & 0.794  & 0.623       & \multicolumn{1}{c}{$\overline{0.340}$} & \multicolumn{1}{c}{\underline{0.333}} & $\mathbf{0.244}$ \\ \hline
		\end{tabular}
	}
	\quad
	\subfloat[Diffraction]{
		\centering
		\begin{tabular}{|c|cccccc|}
			\hline
			\multirow{2}{*}{Var.} & \multirow{2}{*}{$\textrm{GINO}$} & \multirow{2}{*}{$\textrm{GNOT}$} & \multirow{2}{*}{$\textrm{MGN}$} & \multicolumn{3}{c|}{$\textrm{CPGNet}$}                                    \\ \cline{5-7}
			&            &        &                      & \multicolumn{1}{c}{$\mathsf{GAT}$}   & \multicolumn{1}{c}{$\mathsf{GT}$}    & $\mathsf{EConv}$ \\ \hline
			$\rho$       & 0.215  & 0.251  & 0.236       & \multicolumn{1}{c}{$\overline{0.116}$} & \multicolumn{1}{c}{\underline{0.112}} & $\mathbf{0.110}$ \\ 
			$v_1$        & 0.108  & 0.136  & 0.123       & \multicolumn{1}{c}{$\overline{0.060}$} & \multicolumn{1}{c}{\underline{0.059}} & $\mathbf{0.058}$ \\ 
			$v_2$        & 0.044  & 0.058  & 0.052       & \multicolumn{1}{c}{$\overline{0.028}$} & \multicolumn{1}{c}{\underline{0.023}} & $\mathbf{0.025}$ \\
			$p$          & 0.439  & 0.516  & 0.482       & \multicolumn{1}{c}{$\overline{0.239}$} & \multicolumn{1}{c}{\underline{0.235}} & $\mathbf{0.232}$ \\ \hline
		\end{tabular}
	}
	\\
	\subfloat[Supersonic Cylinder]{
		\centering
		\begin{tabular}{|c|cccccc|}
			\hline
			\multirow{2}{*}{Var.} & \multirow{2}{*}{$\textrm{GINO}$} & \multirow{2}{*}{$\textrm{GNOT}$} & \multirow{2}{*}{$\textrm{MGN}$} & \multicolumn{3}{c|}{$\textrm{CPGNet}$}                                     \\ \cline{5-7}
			&            &        &                      & \multicolumn{1}{c}{$\mathsf{GAT}$}   & \multicolumn{1}{c}{$\mathsf{GT}$}    & $\mathsf{EConv}$ \\ \hline
			$\rho$       & 0.094  & 0.103  & 0.059       & \multicolumn{1}{c}{$\overline{0.054}$} & \multicolumn{1}{c}{\underline{0.050}} & $\mathbf{0.047}$ \\ 
			$v_1$        & 0.079  & 0.101  & 0.047       & \multicolumn{1}{c}{$\overline{0.051}$} & \multicolumn{1}{c}{\underline{0.049}} & $\mathbf{0.047}$ \\ 
			$v_2$        & 0.072  & 0.095  & 0.052       & \multicolumn{1}{c}{$\overline{0.050}$} & \multicolumn{1}{c}{\underline{0.048}} & $\mathbf{0.046}$ \\ 
			$p$          & 0.108  & 0.122  & 0.067       & \multicolumn{1}{c}{\underline{0.056}} & \multicolumn{1}{c}{$\overline{0.057}$} & $\mathbf{0.053}$ \\ \hline
		\end{tabular}
	}
	
\end{table}

\begin{figure}[!htb]
	\centering
	\subfloat[Supersonic Bump (second-order DGSEM)]{
		\includegraphics[scale=0.27]{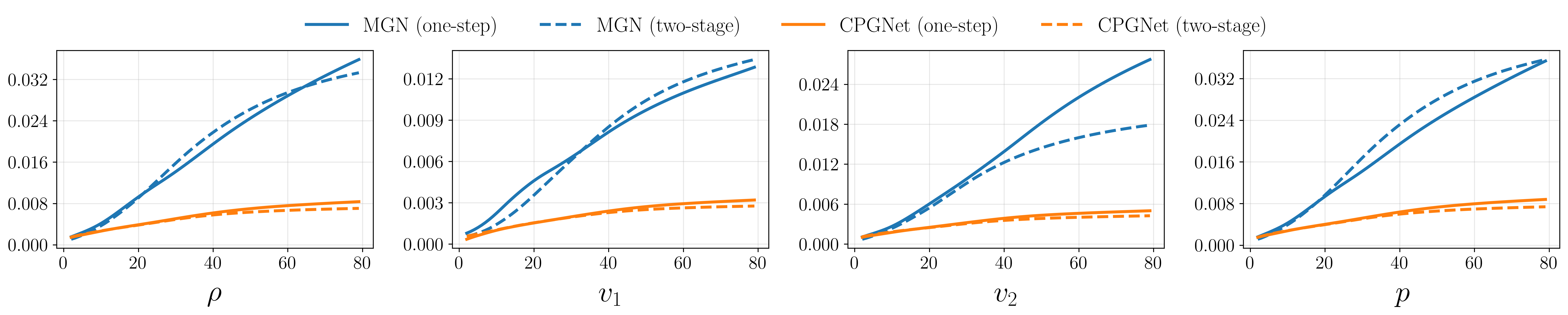}}
	\\
	\subfloat[Supersonic Bump]{
		\includegraphics[scale=0.27]{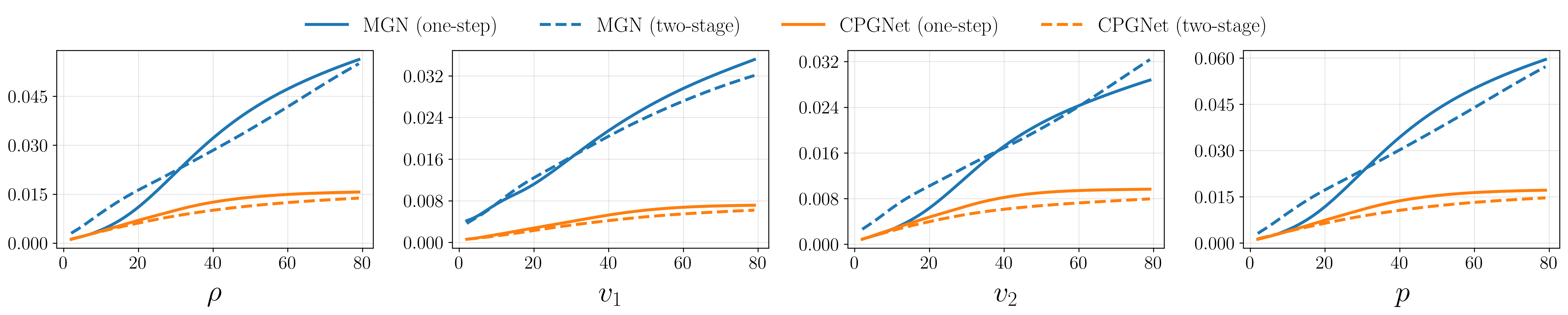}}
	\\
	\subfloat[Forward Step]{
		\includegraphics[scale=0.27]{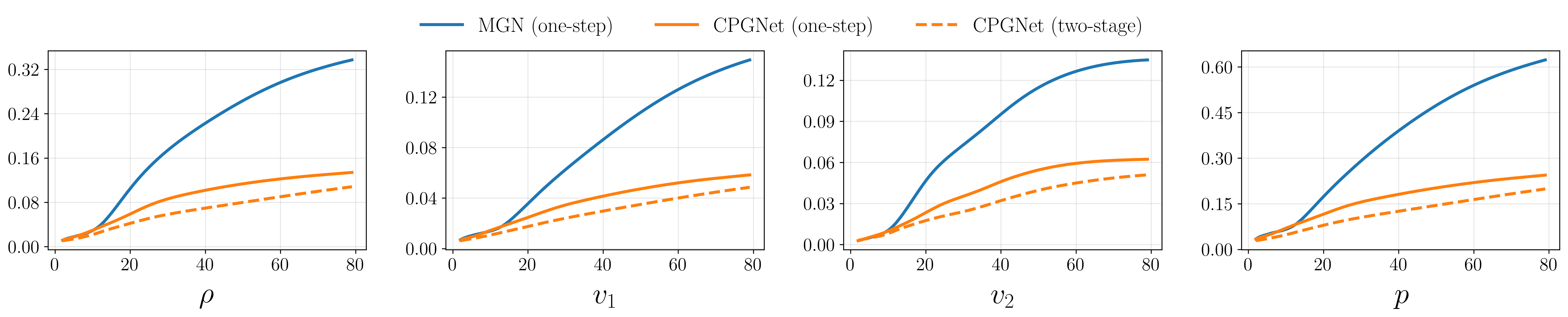}}
	\\
	\subfloat[Diffraction]{
		\includegraphics[scale=0.27]{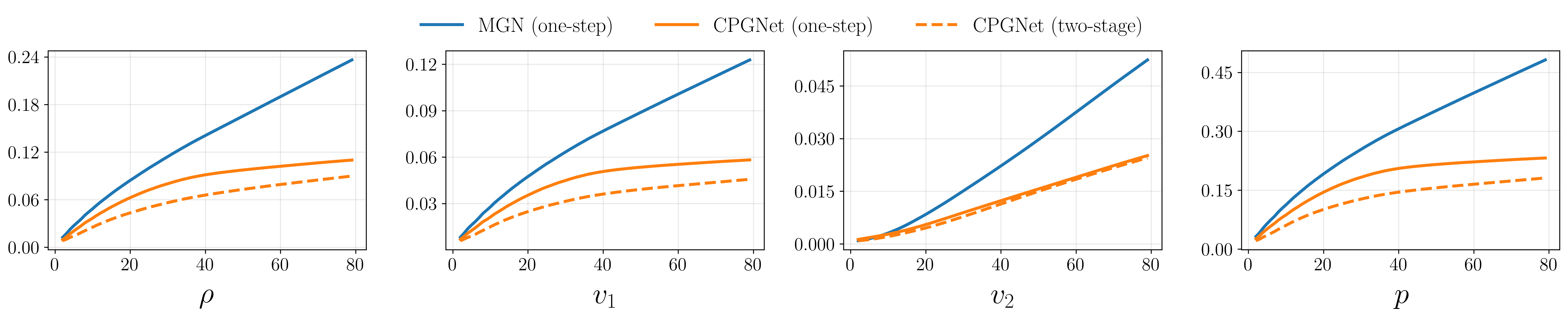}}
	\\
	\subfloat[Supersonic Cylinder]{
		\includegraphics[scale=0.27]{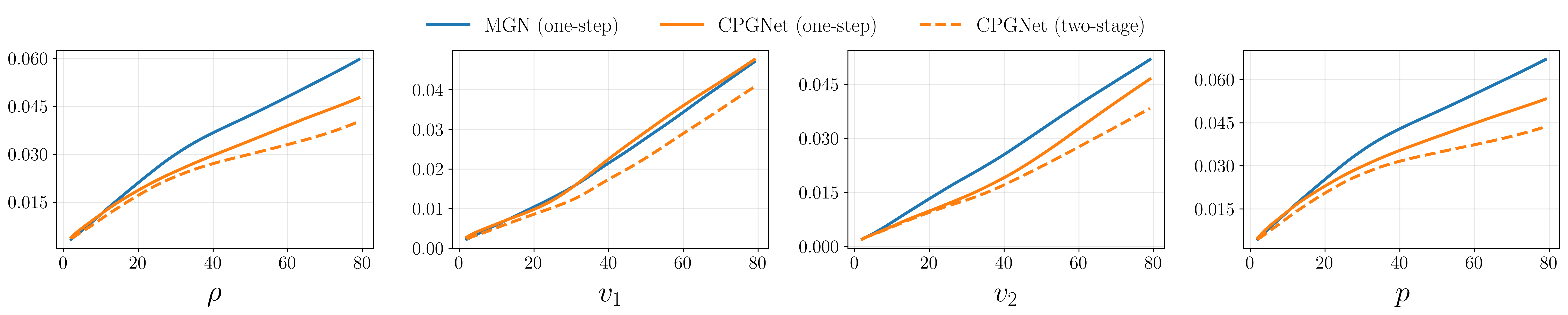}}
	\caption{Mean RMSE over all test cases vs. time steps under the two training strategies.}
	\label{fig:EvolutionsRollout}
\end{figure}

Table~\ref{tab:RMSE_1} summarizes the results under the one-step training. For the processor module of CPGNet, we further compare the enriched edge convolution $\mathsf{EConv}$ with two widely-used attention-based GNNs: the graph attention operator ($\mathsf{GAT}$) and the graph transformer ($\mathsf{GT}$), as detailed in Section~\ref{AppenA}. Across all datasets and variables, our graph neural solver consistently achieves the lowest rollout errors $\text{RMSE}_{\text{all}}$, substantially outperforming the baselines. In particular, CPGNet equipped with $\mathsf{EConv}$ delivers the best performance in nearly all cases, reducing errors by approximately 40\%–80\%, depending on the dataset and variable. Comparisons among different processor variants indicate that the attention-based operators ($\mathsf{GAT}$ and $\mathsf{GT}$) do not offer advantages over $\mathsf{EConv}$, suggesting that the edge-enriched message passing is more effective for capturing local nonlinear interactions of hyperbolic PDEs. Although RMSE is an absolute, scale-dependent metric whose magnitude varies across flow variables, the relative performance trends remain consistent throughout the tests. These results confirm the superior accuracy and rollout robustness of the proposed solver framework.

To assess how solution errors accumulate during rollouts, \textbf{Figure~\ref{fig:EvolutionsRollout}} shows the evolution of the mean RMSE over successive time steps, averaged across all test cases under the two training strategies. While the MGN baseline displays pronounced error growth as the rollout proceeds—ultimately degrading long-term predictive accuracy—the proposed CPGNet‑based solver effectively suppresses temporal error propagation, maintaining a markedly more controlled and stable error profile. Furthermore, adopting the two‑stage training strategy yields additional gains, highlighting the benefit of a multistep loss in capturing temporal dependencies and boosting the reliability of autoregressive inference.

\subsection{Qualitative results}

To complement the quantitative assessment, we plot flow fields at the final time step for representative test trajectories. The proposed graph neural solver remains stable in regions of intense nonlinear wave interactions, producing high‑fidelity rollouts across diverse supersonic regimes. Visual comparisons confirm that our method is markedly superior to the MGN baseline and generalizes robustly to varying geometries, Mach numbers, and flow complexities.

\subsubsection{Supersonic Bump}

\begin{figure}[!b]
	\centering
	\subfloat[$M_\infty = 1.92$]{
		\includegraphics[scale=0.34]{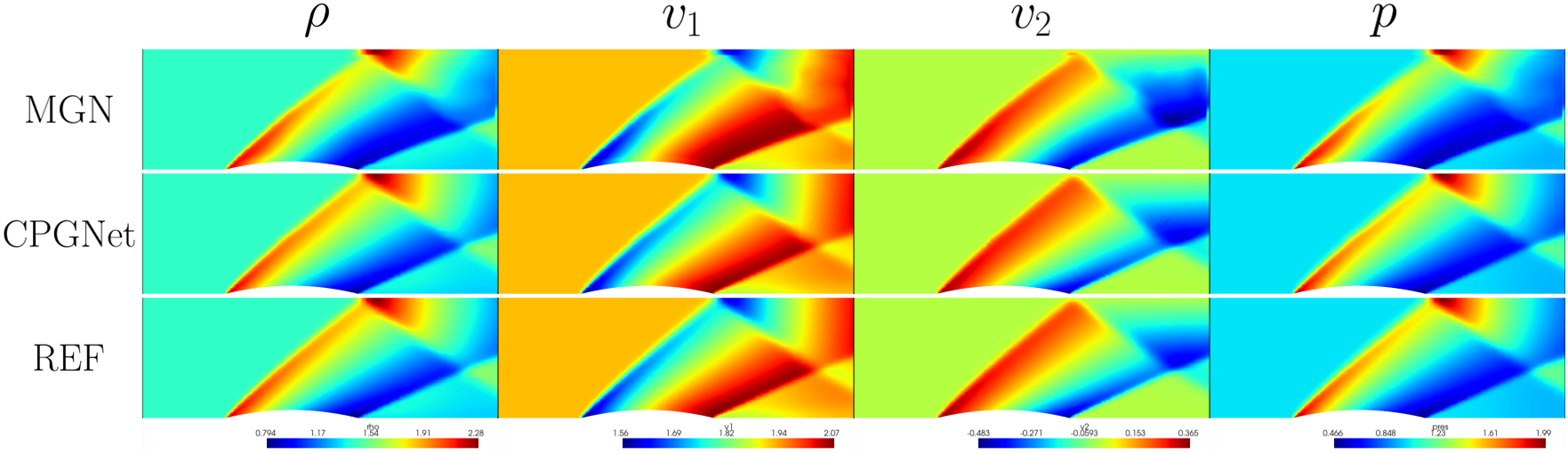}}
	\vspace{-3pt}
	\subfloat[$M_\infty = 1.93$]{
		\includegraphics[scale=0.34]{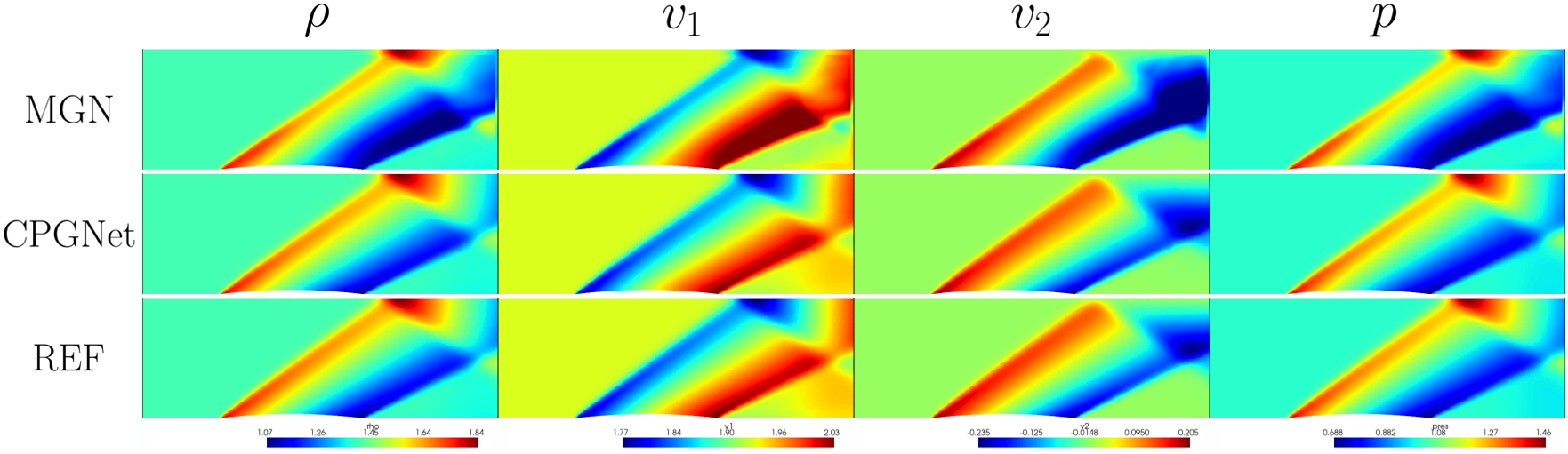}}
	
	\caption{Solution fields of two representative cases in the Supersonic Bump dataset (generated with the second-order DGSEM).}
	\label{fig:Bump1st_Profs}
	
	\vspace{9pt}
	
	\includegraphics[scale=0.51]{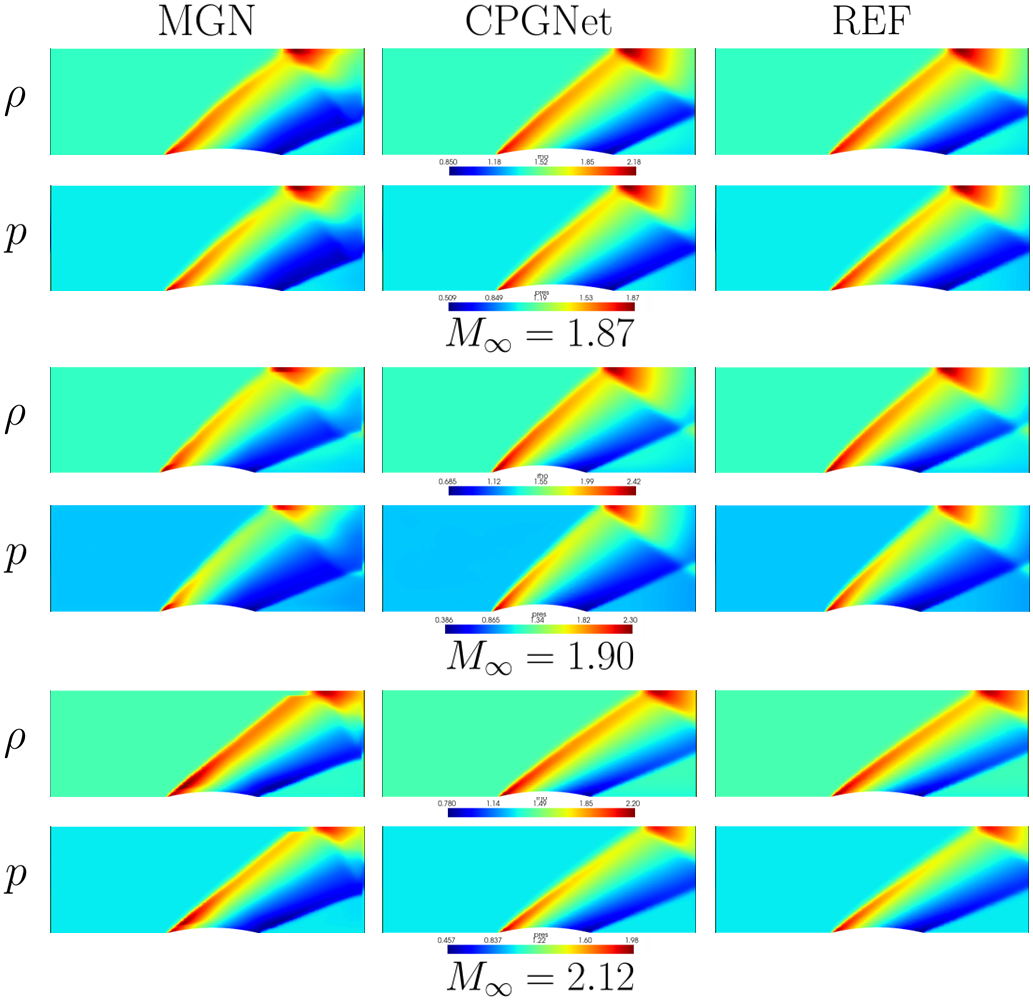}
	\captionof{figure}{Density and pressure fields of representative cases in the Supersonic Bump dataset (generated with the second-order DGSEM).}
	\label{fig:Bump1st_SubPlots}
\end{figure}

We begin by visualizing the final‑time flow fields for two representative test cases from the supersonic‑bump dataset generated with the second‑order DGSEM ($K=1$). \textbf{Figure~\ref{fig:Bump1st_Profs}} compares the high‑fidelity reference (REF) solution with the predictions of the neural solvers; \textbf{Figure~\ref{fig:Bump1st_SubPlots}} further shows density and pressure fields for three additional cases. This flow is characterized by oblique shocks originating at the bump corners. The leading shock reflects off the upper wall and interacts with the trailing shock, producing complex shock–shock and shock–wall interference patterns. Across all cases, CPGNet faithfully captures the global shock topology and reproduces the interaction patterns seen in the reference. In particular, near the shock apex, the predicted shock location, inclination, and thickness closely match the reference, retaining a sharply resolved profile. By contrast, the MGN baseline exhibits pronounced shock smearing and positional drift, reflecting excessive numerical dissipation accumulated during the rollout. 

\begin{figure}[!b]
	\centering
	\subfloat[$M_\infty = 1.92$]{
		\includegraphics[scale=0.34]{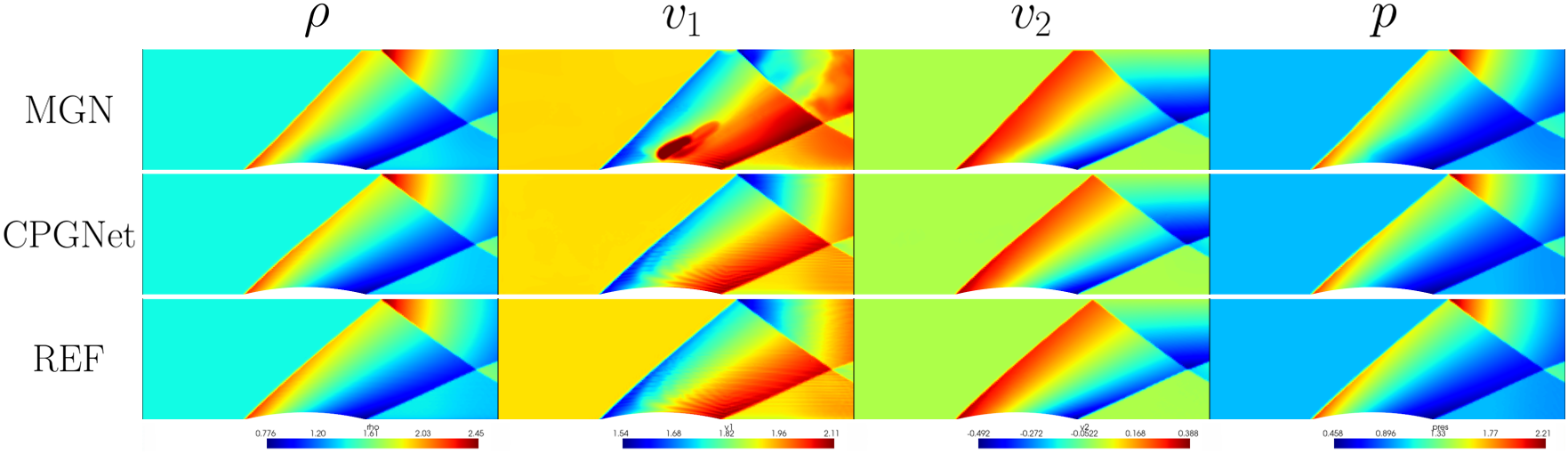}}
	\vspace{-3pt}
	\subfloat[$M_\infty = 2.18$]{
		\includegraphics[scale=0.34]{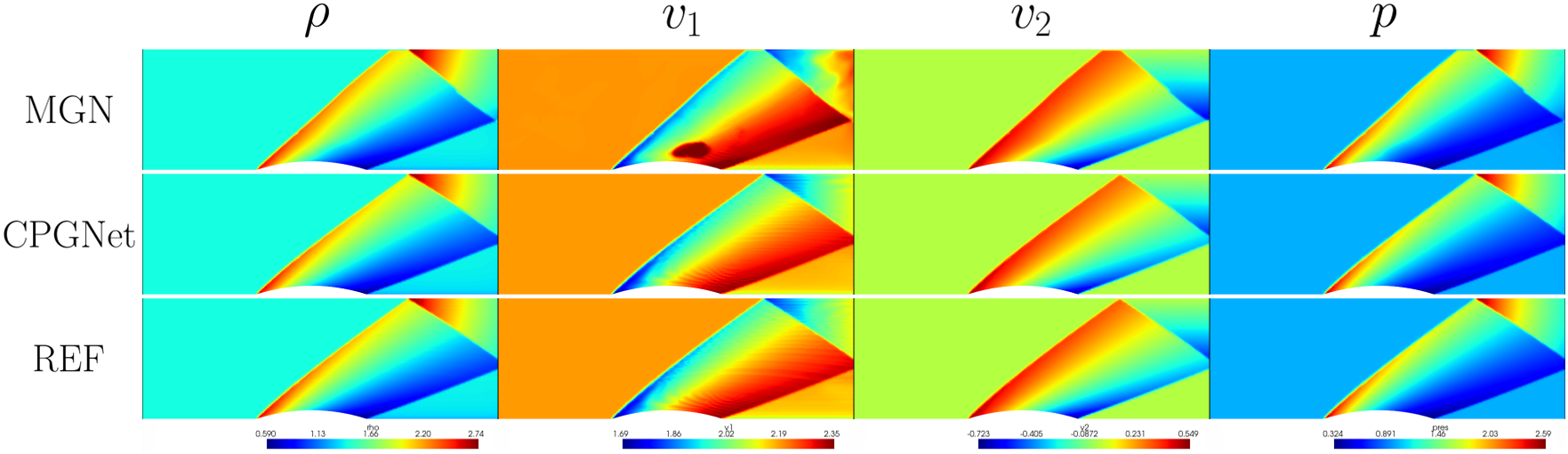}}
	
	\caption{Solution fields of two representative cases in the Supersonic Bump dataset (generated with the fourth-order DGSEM).}
	\label{fig:Bump_Profs}
	
	\vspace{9pt}
	
	\includegraphics[scale=0.51]{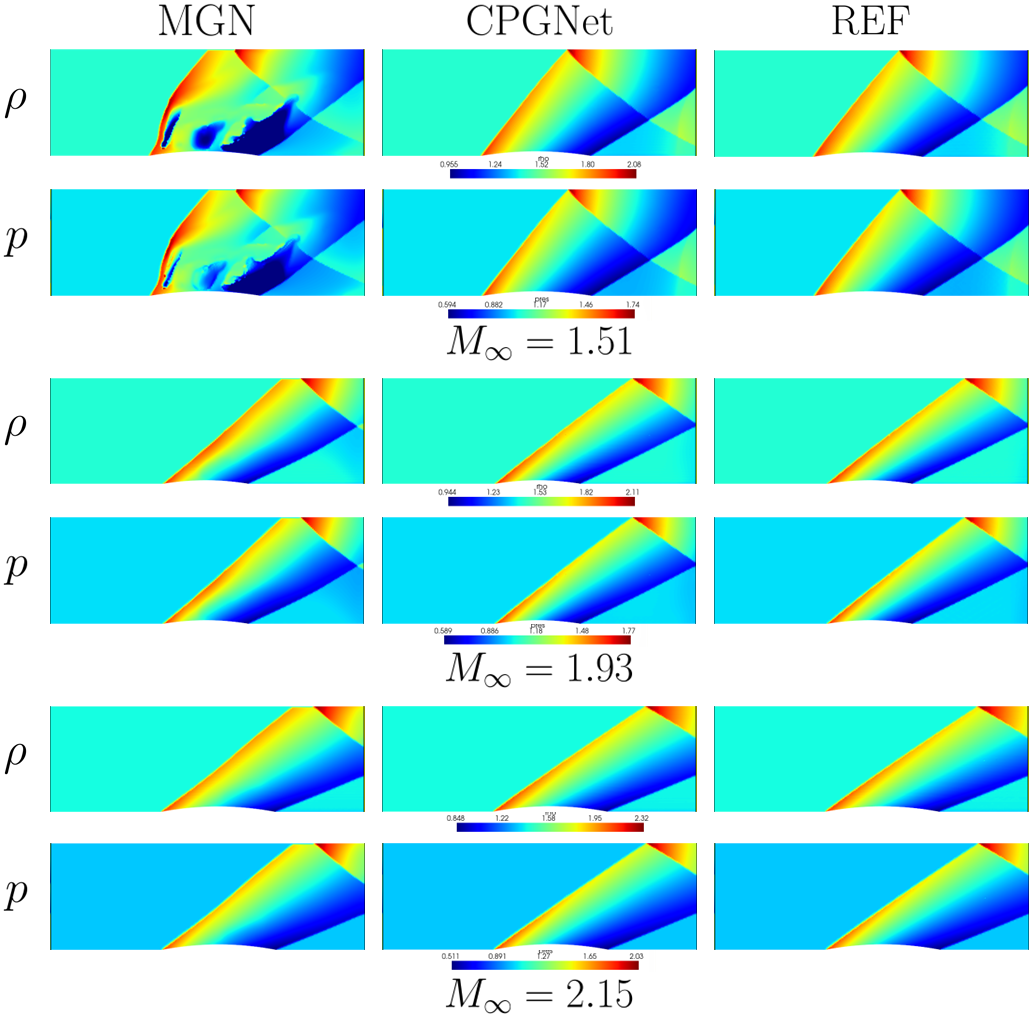}
	\captionof{figure}{Density and pressure fields of representative cases in the Supersonic Bump dataset (generated with the fourth-order DGSEM).}
	\label{fig:Bump_SubPlots}
\end{figure}

CPGNet also delivers superior accuracy in smooth but dynamically sensitive regions. In the stagnation zone upstream of the bump leading edge, where the flow transitions to subsonic and pressure gradients are pronounced, CPGNet restores the reference profiles. MGN predictions are overly smoothed and exhibit distorted contours, reflecting shock propagation errors and insufficient preservation of local solution structure. These observations highlight that CPGNet not only enhances shock capturing but also ensures the fidelity of the coupled shock–subsonic flow field. To further assess performance in smoother flow regimes, \textbf{Figure~\ref{fig:Bump_Profs}} shows final solution fields of two representative test cases from the dataset generated using the fourth-order DGSEM ($K=3$), and \textbf{Figure~\ref{fig:Bump_SubPlots}} provides comparisons of density and pressure for three additional cases. As can be seen, CPGNet maintains excellent agreement with the high-order reference across both shock-dominated and smooth regions, preserving sharp features with minimal degradation. MGN, on the other hand, shows clear visual artifacts and pattern deviations, consistent with the large rollout errors reported in Table~\ref{tab:RMSE_1}.

\FloatBarrier

\FloatBarrier

\begin{figure}[!b]
	\centering
	\subfloat[$M_\infty = 2.77$]{
		\includegraphics[scale=0.34]{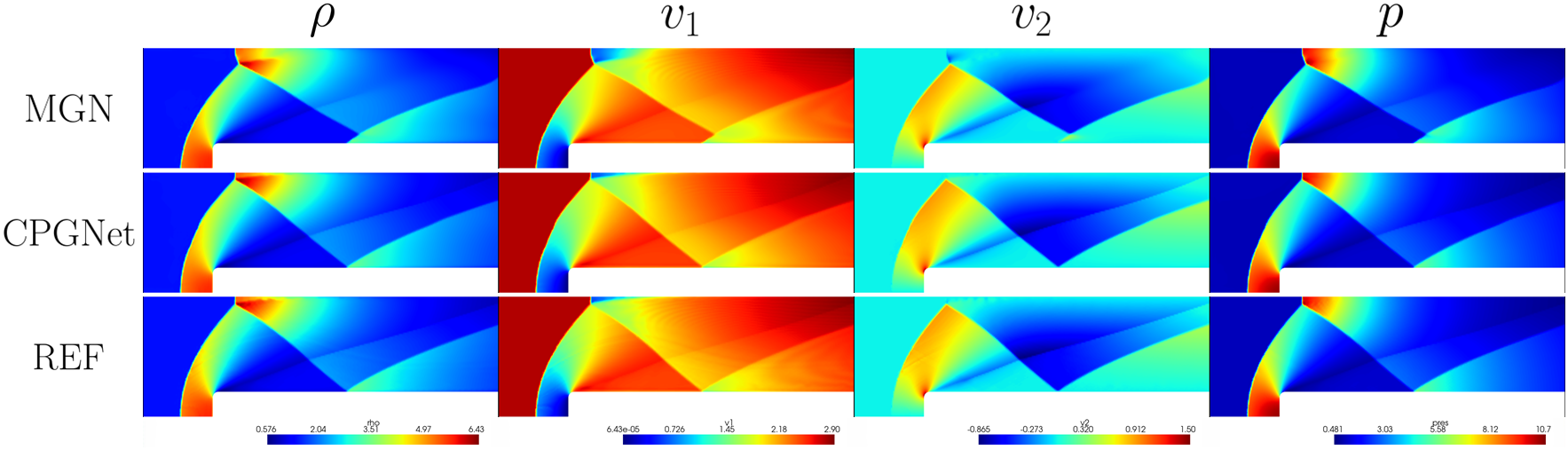}}
	\vspace{-3pt}
	\subfloat[$M_\infty = 2.89$]{
		\includegraphics[scale=0.34]{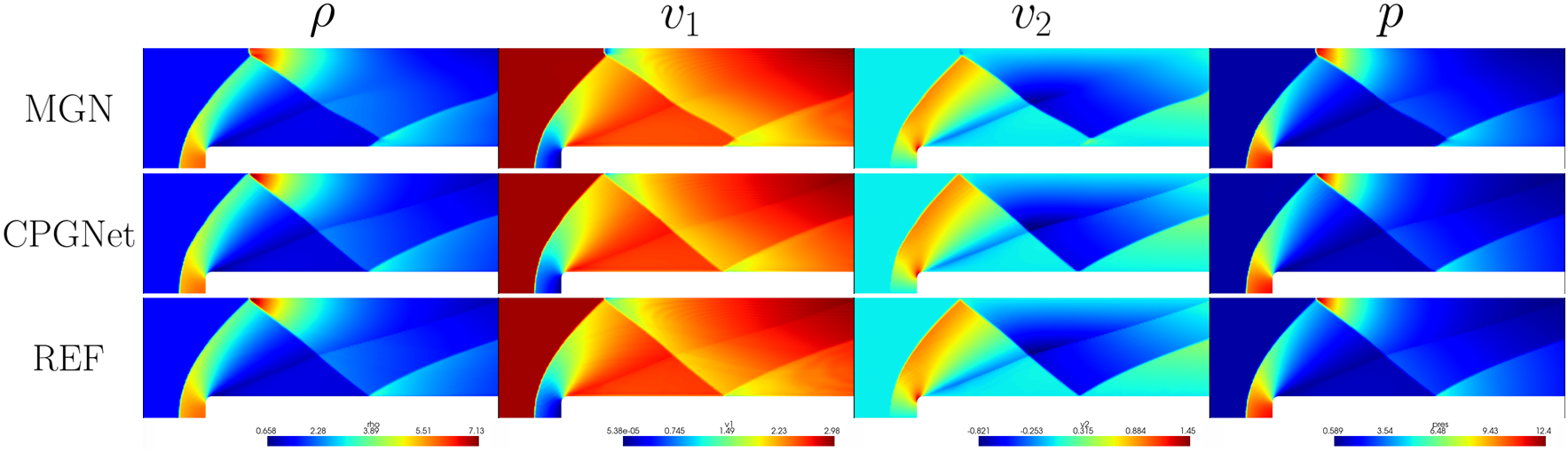}}
	
	\caption{Solution fields of two representative cases in the Forward Step dataset.}
	\label{fig:forward_Profs}
	
	\vspace{9pt}
	
	\includegraphics[scale=0.57]{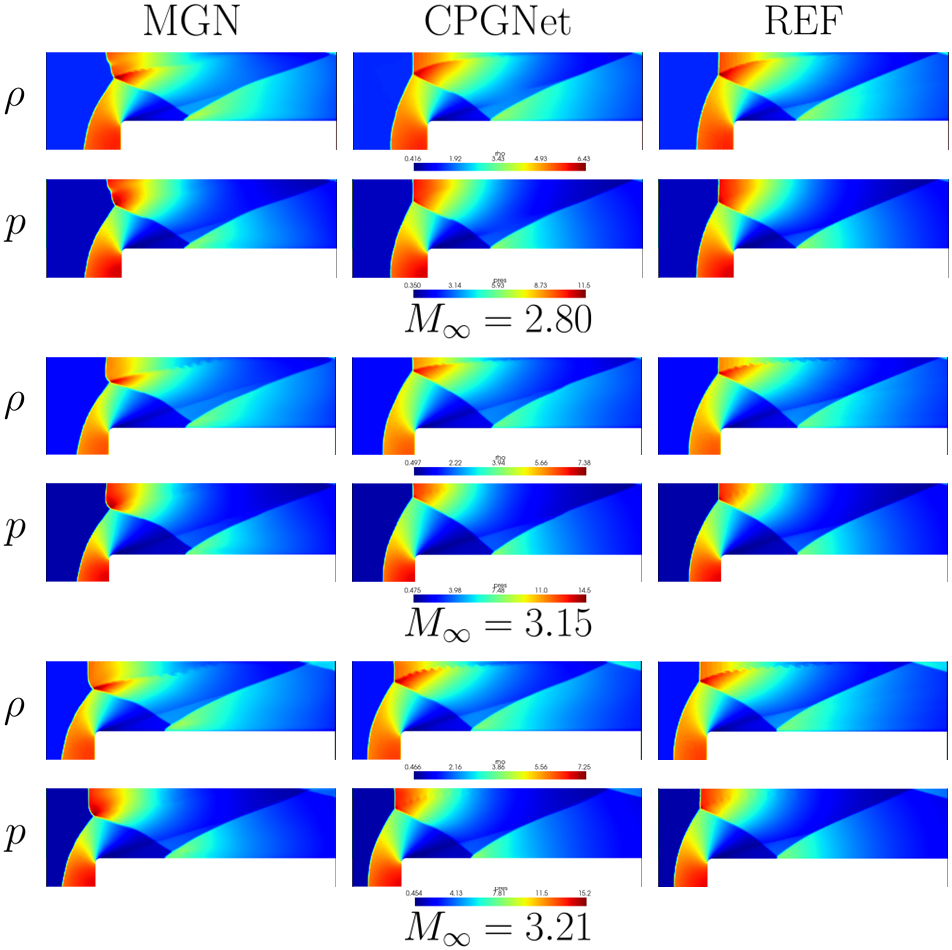}
	\caption{Density and pressure fields of representative cases in the Forward Step dataset.}
	\label{fig:forward_SubPlots}
\end{figure}

\FloatBarrier

\subsubsection{Forward-facing Step}

This benchmark provides a demanding test for shock‑capturing schemes, requiring both the resolution of fine‑scale vortical structures and the suppression of spurious numerical oscillations near strong shocks. Solution fields of representative test cases from the dataset generated with the fourth-order DGSEM are shown in \textbf{Figure~\ref{fig:forward_Profs}} and \textbf{Figure~\ref{fig:forward_SubPlots}}. While the MGN baseline recovers the primary shock topology, it reveals key limitations in handling complex wave interactions. Although it captures the flow expansion at the step corner, the method exhibits severe oscillations around the critical triple point formed during Mach reflection—a known challenge for classical numerical schemes—and yields a smeared representation of the slip line that emanates from the triple point, thereby blurring the entropy‑distinct flow regions.

In contrast, the CPGNet‑based solver effectively suppresses such oscillations not only near the triple point but throughout the computational domain. Moreover, CPGNet robustly captures the downstream flow evolution, including the transition from an initially detached shock to a lambda shock, its reflection from the upper wall, and subsequent wave interactions. Key features—contact discontinuities, multiple shock reflections, and well-defined slip lines—are preserved without introducing non‑physical artifacts, showcasing a clear advance in numerical stability and predictive fidelity.

\begin{figure}[!htb]
	\centering
	\includegraphics[scale=0.36]{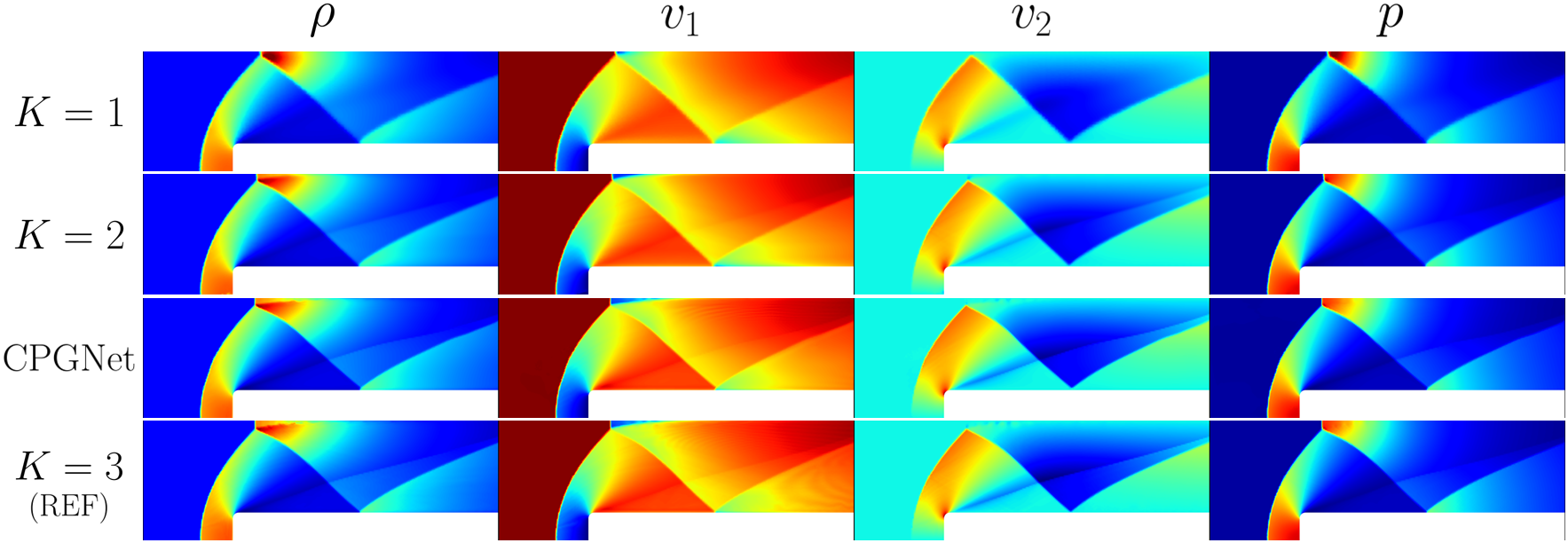}
	\caption{Comparison of solution fields across DGSEM polynomial degrees and graph neural solver for a representative case in the Forward Step dataset.}
	\label{fig:forward_OrderCompare}
\end{figure}

To further examine the effect of numerical order, solution fields of a representative case are compared across DGSEM discretizations with different polynomial degrees, using the $K=3$ simulation as the high-resolution reference. Corresponding predictions from the graph neural solver (CPGNet) are included for direct comparison. As visualized in \textbf{Figure~\ref{fig:forward_OrderCompare}}, lower-order DGSEM solutions exhibit increased numerical diffusion and reduced resolution of shock structures, whereas CPGNet closely reproduces the key flow features of the reference, including sharp shock fronts and post-shock flow patterns.

\begin{table}[!htb]
	\centering
	\caption{Computational time (s) required to simulate a full trajectory for the representative Forward Step case using CPGNet inference and DGSEM with different polynomial degrees.}
	\label{tab:ComputationalEfficiency}
	\begin{tabular}{|c|c|c|c|}
		\hline
		$\textrm{CPGNet}$ & $K=1$  & $K=2$   & $K=3$   \\ \hline
		2.0    & 33.2 & 135.6 & 457.5 \\ \hline
	\end{tabular}
\end{table}

Our graph neural solver delivers a substantial computational advantage over traditional high‑order DGSEM simulations, as quantified in Table~\ref{tab:ComputationalEfficiency}. For the representative case, CPGNet completes a full trajectory in roughly two seconds, achieving speed‑up factors exceeding $10\times$ relative to the lowest‑order DGSEM and over $100\times$ compared to the high‑resolution ($K=3$) reference. All DGSEM runs were performed on a 4th‑generation Intel Xeon CPU using shared-memory multi-threaded parallelization. Importantly, these efficiency gains are attained while CPGNet preserves solution fidelity on par with the $K=3$ DGSEM profiles. The combination of high accuracy and orders‑of‑magnitude faster inference underscores the potential of the proposed framework for real‑time prediction and large‑scale parametric analysis of compressible flows.

\subsubsection{Shock Diffraction}

\begin{figure}[!t]
	\centering
	\subfloat[$M_\infty = 2.72$]{
		\includegraphics[scale=0.29]{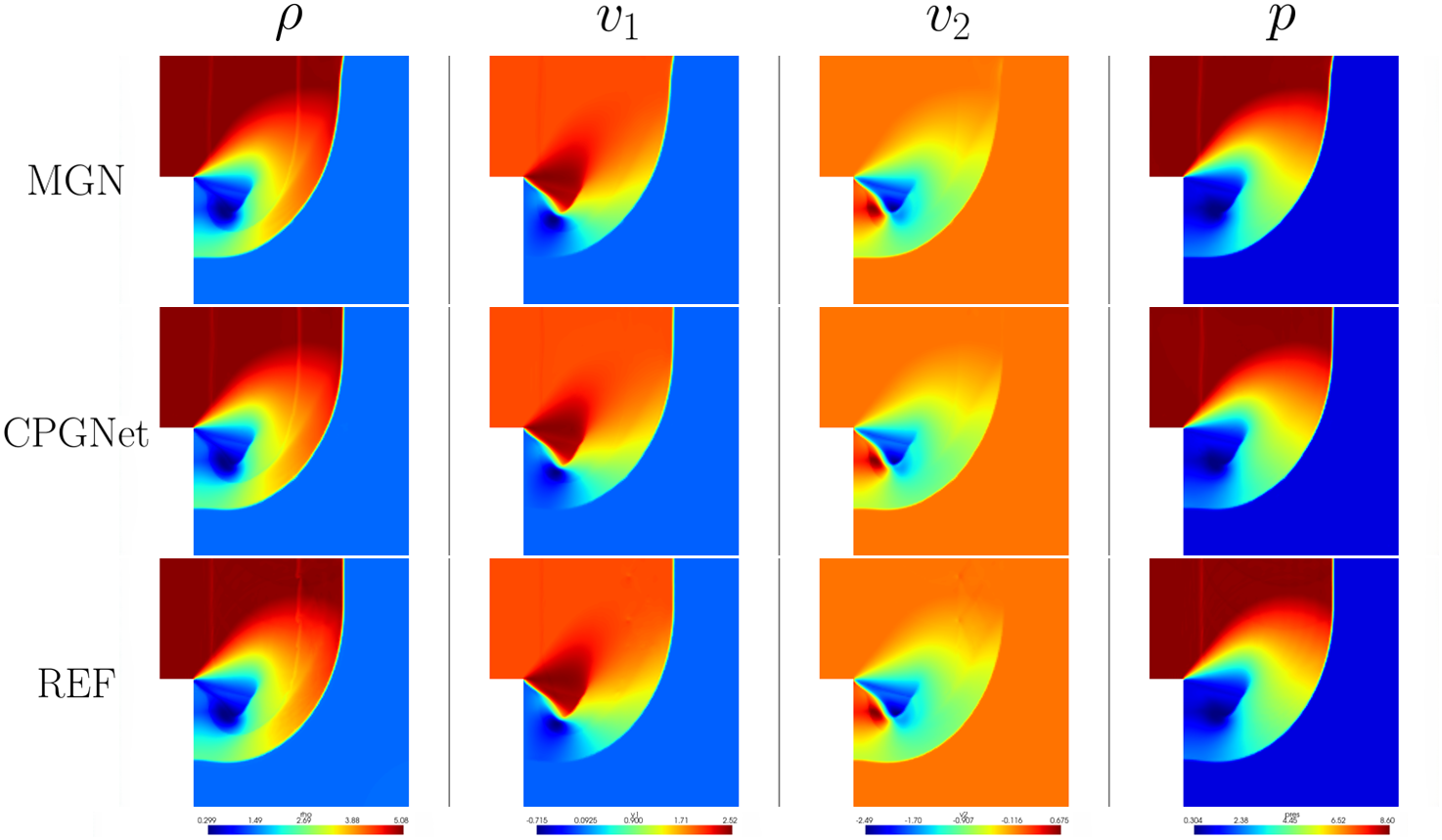}}
	\vspace{-3pt}
	\subfloat[$M_\infty = 3.28$]{
		\includegraphics[scale=0.29]{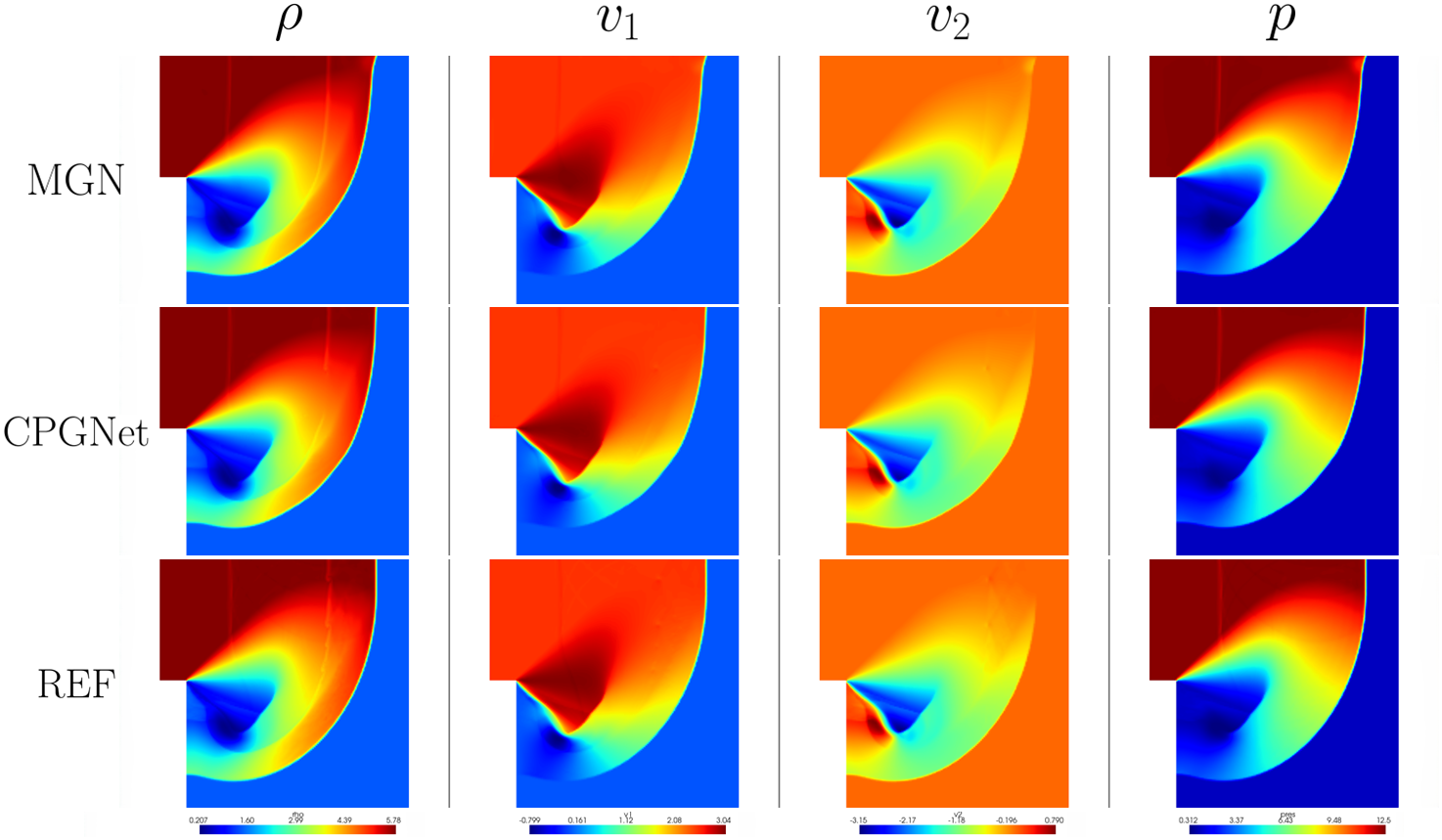}}
	
	\caption{Solution fields of two representative cases in the Diffraction dataset.}
	\label{fig:diffraction_Profs}
	
	\vspace{9pt}
	
	\includegraphics[scale=0.44]{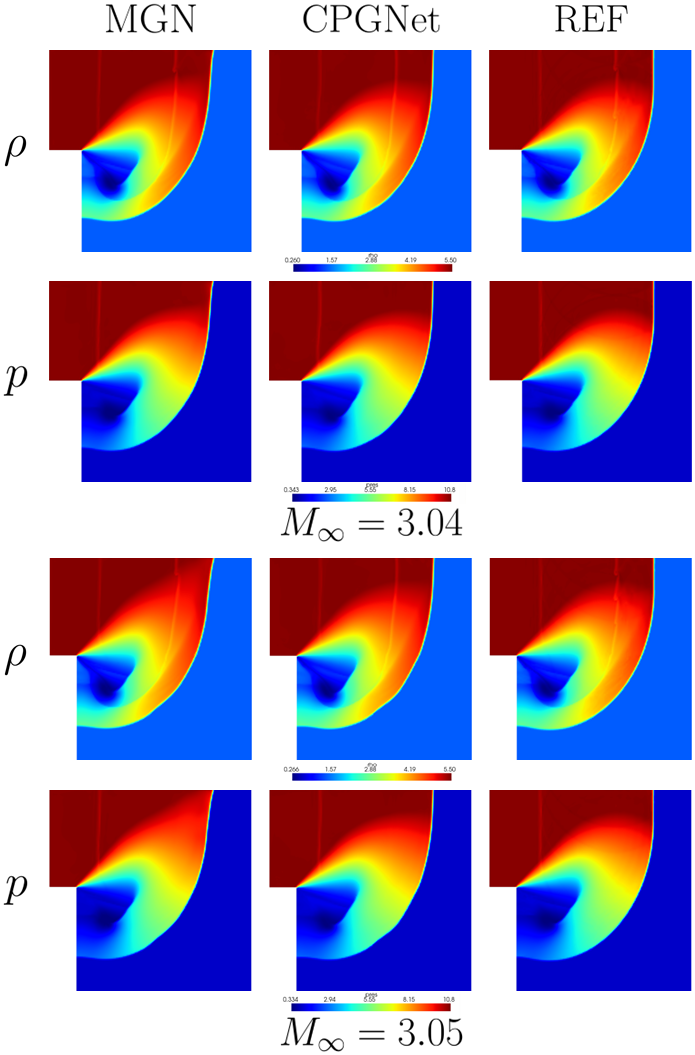}
	\caption{Density and pressure fields of representative cases in the Diffraction dataset.}
	\label{fig:diffraction_SubPlots}
\end{figure}

\textbf{Figure~\ref{fig:diffraction_Profs}} and \textbf{Figure~\ref{fig:diffraction_SubPlots}} present representative results from the Shock Diffraction benchmark, a test problem that showcases complex wave patterns arising from the shock interaction with a sharp corner. While both methods capture the primary flow structures—the diffracted shock, Mach stem, and expansion fan—clear differences emerge near the upper boundary of the expansion fan. The MGN baseline shows local contour distortions and mild non‑physical oscillations in this region, whereas CPGNet maintains physically coherent contours that align with the reference. Across all tested cases, CPGNet demonstrates consistently higher stability and visual fidelity when resolving strong shock–expansion interactions.

\subsubsection{Supersonic Cylinder}

\begin{figure}[!t]
	\centering
	\subfloat[$M_\infty = 1.40$]{
		\includegraphics[scale=0.3]{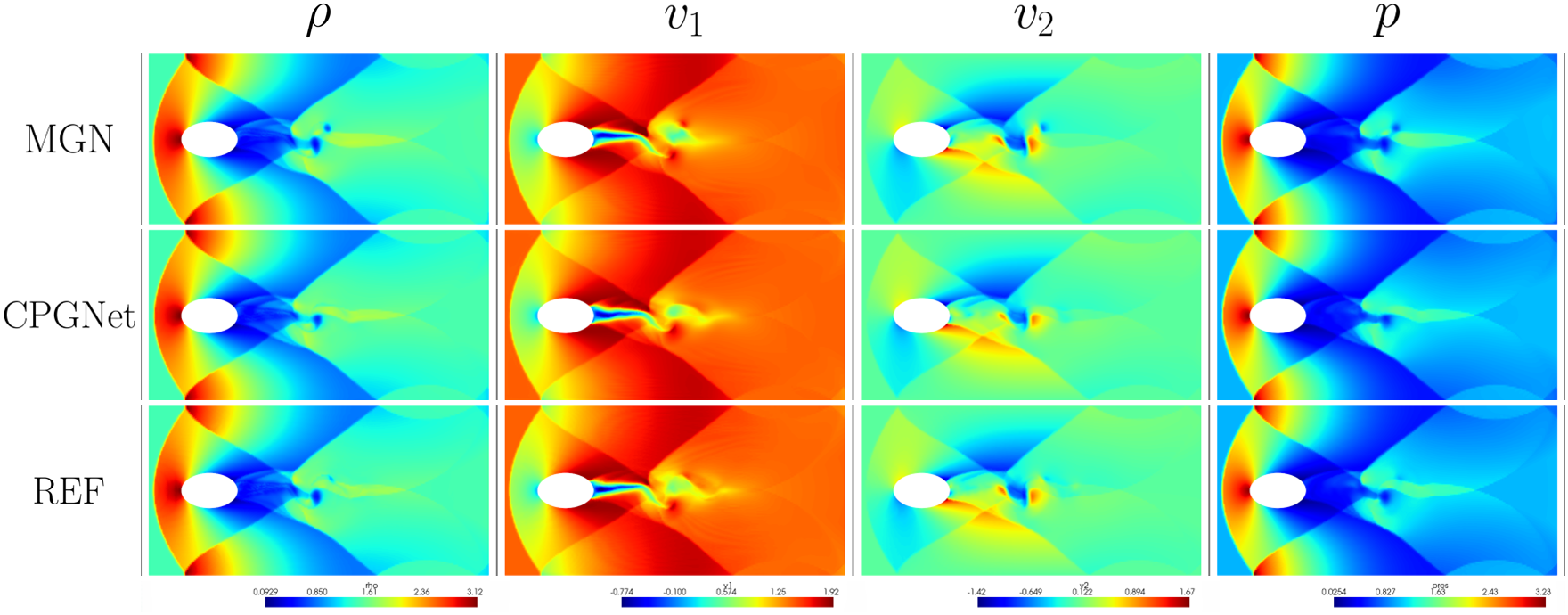}}
	\vspace{-3pt}
	\subfloat[$M_\infty = 1.71$]{
		\includegraphics[scale=0.3]{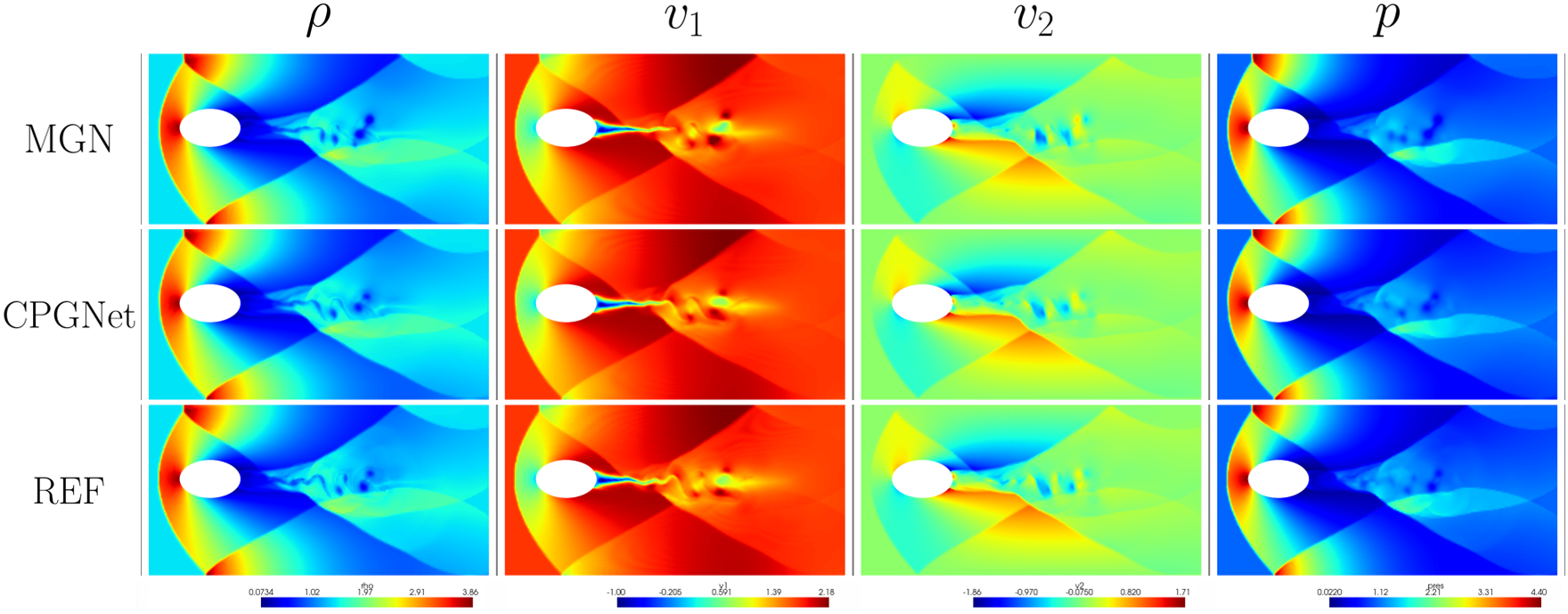}}
	
	\caption{Solution fields of two representative cases in the Supersonic Cylinder dataset.}
	\label{fig:Cylinder_Profs}
	
	\vspace{9pt}
	
	\includegraphics[scale=0.53]{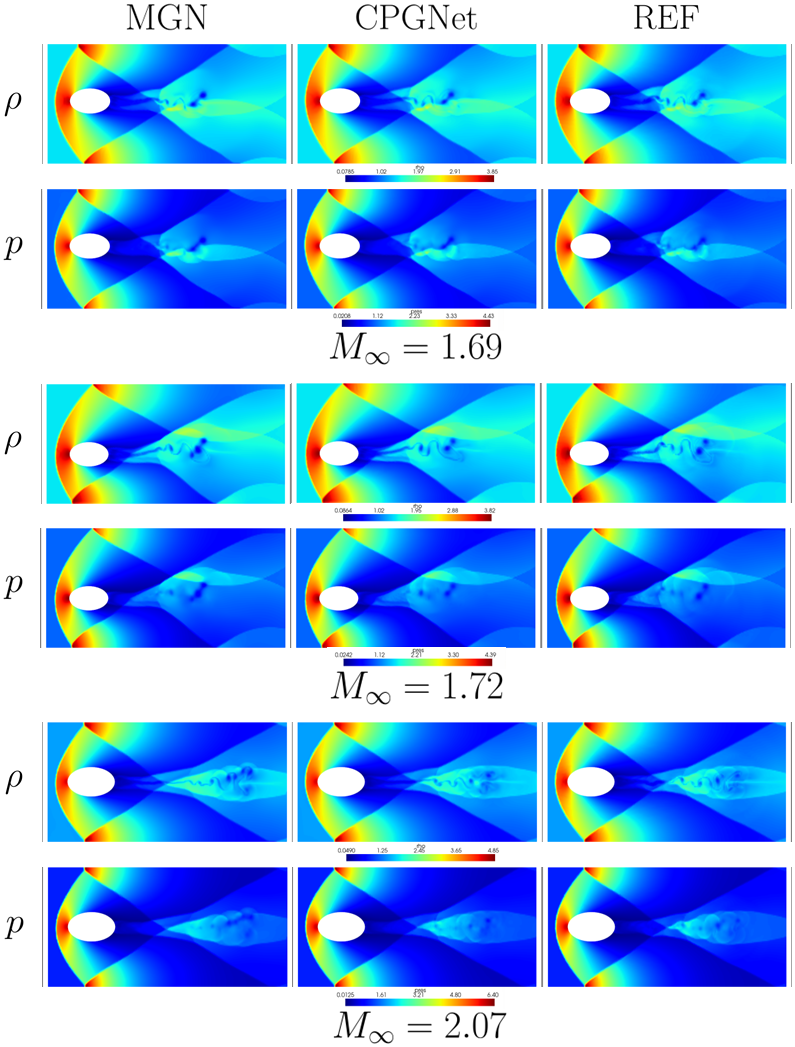}
	\caption{Density and pressure fields of representative cases in the Supersonic Cylinder dataset.}
	\label{fig:Cylinder_SubPlots}
\end{figure}

\textbf{Figure~\ref{fig:Cylinder_Profs}} and \textbf{Figure~\ref{fig:Cylinder_SubPlots}} show comparisons for the supersonic flow past an elliptic cylinder, a configuration characterized by a detached bow shock upstream, strong flow acceleration around the body, and a complex, unsteady separated wake with vortex shedding downstream. The MGN baseline captures the primary features, including the global bow‑shock structure and the expansion fans. However, its prediction of the downstream separated wake appears overly diffused and geometrically distorted. By comparison, the proposed CPGNet solver yields smaller discrepancies in the shape and details of the wake. The formation of a well-defined Mach disk and the associated triple points—where the incident shock, reflected shock, and Mach stem meet—are also accurately resolved, together with the emanating slip lines that separate regions of different entropy.

Overall, the qualitative comparisons confirm that the structure‑preserving design of CPGNet delivers stable and accurate spatio‑temporal solutions across all benchmark problems considered. The solver consistently captures global shock topology, fine‑scale discontinuities, and coupled shock–subsonic interactions with high fidelity, while mitigating the non‑physical oscillations and excessive dissipation prevalent in the MGN baseline. Its robust performance generalizes across a broad spectrum of Mach numbers, geometries, and flow regimes, demonstrating strong adaptability to unseen configurations. These findings mark a distinct step forward relative to black‑box neural surrogates and underscore the reliability of the CPGNet framework for challenging supersonic flows.

\FloatBarrier

\FloatBarrier

\section{Conclusion}

We have presented an interpretable, structure‑preserving graph neural solver designed for parametric hyperbolic conservation laws. At its core, this work bridges the robustness and interpretability of classical Godunov‑type finite‑volume schemes with the flexibility and expressive capacity of graph neural networks (GNNs). We show that successive message‑passing and aggregation in GNNs naturally emulate the one‑sided biasing and wide stencils characteristic of high‑order nonlinear reconstructions. Departing from black‑box neural surrogates, our framework treats the network as a learned reconstruction‑and‑flux operator—decoding interface states from edge representations, coupling them with a differentiable Riemann solver, and advancing the solution via a conservative flux‑aggregation layer. This design enforces discrete local conservation and upwinding directly at the architectural level. To overcome the efficiency limitations imposed by explicit CFL constraints, we further introduce an ADER‑inspired formulation that recasts the message‑passing GNN as a high‑order space‑time predictor. The resulting implicit‑like one‑step update remains fully conservative and Riemann‑solver compatible while operating stably with time steps far exceeding the explicit CFL limit. In addition, learned geometric weights allow the solver to operate directly on point‑cloud simulation data, requiring only graph connectivity and nodal coordinates and eliminating the need for an explicit mesh during inference.

Through systematic validation on four challenging families of supersonic flow benchmarks spanning wide parametric variations, the proposed neural solver consistently outperforms strong baseline methods in both quantitative and qualitative assessments. It achieves markedly improved long‑horizon rollout stability, reduces accumulated errors by up to 80\%, and exhibits superior fidelity in capturing sharp shocks, contact discontinuities, and coupled shock–subsonic interactions. Compared to traditional discontinuous Galerkin spectral element (DGSEM) discretizations, the solver surpasses low‑order variants in accuracy while delivering runtime speedups of more than two orders of magnitude relative to high-resolution reference simulations. These results highlight a viable pathway toward scalable, structure‑aware neural PDE solvers for general conservation laws, with promising implications for large‑scale many‑query workflows such as design optimization and real‑time flow control.

\section{Code Availability}

The source code and datasets are publicly available at
\\
https://gitlab.com/jiaminjiang66/cpggnspdes.

\section{Appendix A. Attention-based Graph Operators} \label{AppenA}

We summarize two widely used attention-based graph operators. Both augment standard message passing by introducing learnable attention weights that modulate the relative contribution of each neighboring node during aggregation.

The Graph Attention (GAT) operator \cite{velivckovic2017graph,brody2021attentive} integrates self attention into graph representation learning, allowing each node to adaptively emphasize the most informative neighbors. For a node pair $(i,j)$, an unnormalized attention score is computed as
\begin{equation}
	\xi_{ij} = \bm{a}^{\top} \mathrm{LeakyReLU}\left( \mathbf{W}_{1} \bm{q}_i + \mathbf{W}_{2} \bm{q}_j \right),
\end{equation}
where $\mathbf{W}$ are learnable weight matrices and $\bm{a}$ is the shared attention vector. The LeakyReLU nonlinearity uses a negative slope of 0.2. These scores are then normalized over the (self-including) neighborhood of node $i$ using a softmax
\begin{equation}
	\psi_{ij} = \mathrm{softmax}_j(\xi_{ij}) = \frac{\mathrm{exp} \left ( \xi_{ij} \right )}{\sum_{k \in\mathcal{N}(i)\cup\left \{ i \right \}} \mathrm{exp} \left ( \xi_{ik} \right ) }.
\end{equation}

The updated embedding is obtained via an attention-weighted combination of transformed node features
\begin{equation}
	{\bm{q}}'_i = \psi_{ii} \mathbf{W}_{1} \bm{q}_i + \! \sum_{j\in\mathcal{N}(i)} \! \psi_{ij} \mathbf{W}_{2} \bm{q}_j
\end{equation}
followed by layer normalization and a ReLU activation. Multi-head attention is realized by constructing multiple attention maps in parallel and then averaging the resulting head-wise outputs.

The Graph Transformer operator \cite{dwivedi2020generalization} instead employs the scaled dot-product attention mechanism popularized by the Transformer architecture \cite{vaswani2017attention}. For each attention head, the neighborhood attention weights are defined by
\begin{equation}
	\psi_{ij} = \mathrm{softmax} \left ( \frac{(\mathbf{W}_{3} \bm{q}_i)^{\top} \left ( \mathbf{W}_{4} \bm{q}_j \right )}{\sqrt{\hbar}} \right ),
\end{equation}
where $\hbar$ denotes the hidden dimension. Message aggregation takes the form
\begin{equation}
	{\bm{q}}'_i = \mathbf{W}_{1} \bm{q}_i + \! \sum_{j\in\mathcal{N}(i)} \! \psi_{ij} \mathbf{W}_{2} \bm{q}_j.
\end{equation}
Note that, in the formulation above, the GT layer remains within the local message-passing setting: attention is restricted to one-hop neighbors, and the learned coefficients $\psi_{ij}$ act solely as adaptive importance weights for aggregating incoming messages.

All network variants follow the Encode-Process-Decode architecture (Section~\ref{basicEPD}). The Processor contains 12 layers and alternates between $\mathsf{EConv}$ and attention-based operators (6 $\mathsf{EConv}$ + 6 $\mathsf{GAT}$/$\mathsf{GT}$). The $\mathsf{EConv}$ blocks propagate and update edge-centric messages, whereas the attention-based blocks refine node-to-node interactions through learned neighbor weights. We use two attention heads and set the hidden feature dimension to 128.

\section{Appendix B. Training Costs} \label{AppenB}

Table~\ref{tab:traincosts} summarizes the wall-clock training costs (hrs) of the MGN surrogate and the graph neural solver for each dataset. For one-step training, CPGNet exhibits training times comparable to, and consistently slightly lower than, those of MGN, indicating no additional overhead from enforcing conservation and upwinding at the architectural level. The two-stage CPGNet strategy introduces a modest additional cost associated with multi-step autoregressive fine-tuning. This additional expense remains moderate relative to the overall training time and is justified by the gains in long-term stability and predictive accuracy.

\begin{table}[!htb]
	\centering
	\caption{Training costs (hrs) of the MGN surrogate and the graph neural solver (CPGNet).}
	\label{tab:traincosts}
	\begin{tabular}{|c|c|cc|}
		\hline
		\multirow{2}{*}{Dataset} & \multirow{2}{*}{\begin{tabular}[c]{@{}c@{}}MGN\\ one-step\end{tabular}} & \multicolumn{2}{c|}{CPGNet}                     \\ \cline{3-4} 
		&                                                                         & \multicolumn{1}{c|}{one-step} & two-stage       \\ \hline
		Supersonic Bump          & 17.7                                                                    & \multicolumn{1}{c|}{16.9}     & 24.3 (16.9+7.4) \\ \hline
		Forward Step             & 20.7                                                                    & \multicolumn{1}{c|}{20.1}     & 28.8 (20.1+8.7) \\ \hline
		Shock Diffraction        & 21.2                                                                    & \multicolumn{1}{c|}{20.6}     & 29.5 (20.6+8.9) \\ \hline
		Supersonic Cylinder      & 22.5                                                                    & \multicolumn{1}{c|}{21.8}     & 31.4 (21.8+9.6) \\ \hline
	\end{tabular}
\end{table}

\section*{Acknowledgements}

This work is partially supported by the NSFC Major Research Plan - Interpretable and General Purpose Next-generation Artificial Intelligence (Nos. 92570001 and 92370205), NSFC grant 12501604, NSFC grant 12425113, the Basic Research Program of Jiangsu under Grant BK20253039 and Grant BK20250475, and Key Laboratory of the Ministry of Education for Mathematical Foundations and Applications of Digital Technology, University of Science and Technology of China.

\bibliography{mybibfile}

\end{document}